# Exploiting offshore wind and solar resources in the Mediterranean using ERA5 reanalysis data

*Takvor H. Soukissian* [a,*], *Flora E. Karathanasi* [a, b], *Dimitrios K. Zaragkas* [c]

[a] Institute of Oceanography, Hellenic Centre for Marine Research, 46.7 km Athens-Sounio Ave., 190 13 Anavyssos, Greece

[b] Cyprus Marine Observation Network, CMMI - Cyprus Marine and Maritime Institute, CMMI House, Vasileos Pavlou Square, P.O. Box 40930, Larnaca 6023, Cyprus

[c] School of Naval Architecture and Marine Engineering, National Technical University of Athens, Iroon Polytechniou 9, Zografos, 157 80 Athens, Greece

[*] Corresponding author:

e-mail: tsouki@hcmr.gr

Phone: 0030-22910-76420

Fax: 0030-22910-76323

**ABSTRACT**

Commercial electricity production from marine renewable sources is becoming a necessity at a global scale. Offshore wind and solar resources can be combined to reduce construction and maintenance costs. In this respect, the aim of this study is two-fold: i) analyse offshore wind and solar resource and their variability in the Mediterranean Sea at the annual and seasonal scales based on the recently published ERA5 reanalysis dataset, and; ii) perform a preliminary assessment of some important features of complementarity, synergy, and availability of the examined resources using an event-based probabilistic approach. A robust coefficient of variation is introduced to examine the variability of each resource and a joint coefficient of variation is implemented for the first time to evaluate the joint variability of offshore wind and solar potential. The association between the resources is examined by introducing a robust measure of correlation, along with the Pearson's $r$ and Kendall's $\tau$ correlation coefficient and the corresponding results are compared. Several metrics are used to examine the degree of complementarity affected by variability and intermittency issues. Areas with high potential and low variability for both resources include the Aegean and Alboran seas, while significant synergy (over 52%) is identified in the gulfs of Lion, Gabes and Sidra, Aegean Sea and northern Cyprus Isl. The advantage of combining these two resources is highlighted at selected locations in terms of the monthly energy production.

**NOMENCLATURE**

**Symbols**

| | |
|---|---|
| $c_0, c_1, c_2, c_3$ | coefficients to estimate PV module temperature |
| $CV$ | coefficient of variation |
| $D_{NW}$ | duration of "no wind" |
| $D_{NS}$ | duration of "no solar" |

| $G, G_{STC}$ | solar irradiance from model and at standard testing conditions |
|---|---|
| $IAV$ | inter-annual variability index |
| $JCV$ | joint coefficient of variation of $WP$ and $SP$ |
| $MAD$ | median absolute deviation |
| $MAV$ | mean annual variability index |
| $MED$ | median |
| $\text{med}(\cdot)$ | median value operator |
| $MV$ | monthly variability index |
| $\rho, r$ | population and sample version of the Pearson correlation coefficient |
| $P_S, P_W$ | solar and offshore wind power output |
| $P_{STC}$ | rated power of the PV module |
| $r_{CMED}$ | median correlation coefficient |
| $RCV$ | robust coefficient of variation |
| $SCW$ | solar to wind complementarity index |
| $SP$ | surface solar radiation downwards |
| $SV$ | seasonal variability index |
| $SWS$ | synergy of wind and solar index |
| $T_{mod}, T_{STC}$ | PV module temperature and reference temperature |
| $u$ | wind speed at 100 m above sea level |
| $u_{10}$ | wind speed at 10 m above sea level |
| $UWS$ | joint non-availability of wind and solar index |
| $WCS$ | wind to solar complementarity index |
| $WP$ | offshore wind power density |
| $\bar{X}$ | sample mean value of variable $X$ |
| $\eta$ | performance factor of PV system |
| $\rho$ | air density |

| $\sigma_S$ | standard deviation of offshore solar irradiance |
|---|---|
| $\sigma_W$ | standard deviation of offshore wind power density |
| $\sigma_{XY}, s_{XY}$ | population and sample covariance of random variables $X$ and $Y$ |
| $\tau, \hat{\tau}$ | population and sample estimate of Kendall correlation coefficient |

**Abbreviations**

| ASCAT | advanced scatterometer |
|---|---|
| CC | correlation coefficient |
| cdf | cumulative distribution function |
| MERRA | Modern-Era Retrospective analysis for Research and Applications |
| ECMWF | European Centre for Medium-range Weather Forecasts |
| ISCCP | International Satellite Cloud Climatology Project |
| IAV | Inter-annual variability |
| MAV | mean annual variability |
| MV | monthly variability |
| NOAA | National Oceanic and Atmospheric Administration |
| OWF | offshore wind farm |
| PV | photovoltaic |
| QuikSCAT | Quick Scatterometer |
| STC | standard testing conditions |
| SV | seasonal variability |
| WRF-ARW | Weather Research and Forecasting - Advanced Research |

1. INTRODUCTION

Both wind and solar energy are renewable energy sources, alternative to fossil fuels, with a long history, the growth of which was significantly boosted by the oil crisis in the 1970s and further instigated by the Kyoto Protocol, adopted in 1997. Solar energy industry began to grow with the development of a silicon-based solar cell by three American scientists working for the Bell Laboratories in the 1950s with a low conversion efficiency (9% in 1958) and a high cost (around 300 US$/W in 1956), [1]. On the other hand, the first wind farm was realized almost 100 years later in the opposite part of the North Atlantic Ocean, in 1980 in New Hampshire, consisting of 20 wind turbines with rated power 0.6 MW.

In recent years, the cost of acquiring land recourses, along with the limited available spaces for new large scale wind and photovoltaic (PV) installations combined with the EU commitments by member states to promote renewable energies for the combat against climate change, turned the interest to the marine environment.

1.1 Offshore wind and solar energy

Offshore winds produce a considerably higher power than onshore, characterized also by less variability. Few studies have been conducted solely in the Mediterranean Sea for the assessment of offshore wind climate and offshore wind power potential at basin level. These studies used wind data derived from different sources. Specifically, in [2], the offshore wind climatology from 6-hour wind speed data obtained from the European Centre for Medium-range Weather Forecasts (ECMWF) for a 24-year period was presented, while in [3], a high-resolution atmospheric hindcast over the Mediterranean Sea using the Weather Research and Forecasting - Advanced Research (WRF-ARW) model was produced. In [4], simulation fields from four regional climate models (Regional Climate Model Version 3/RegCM3, PROTHEUS, and two versions of WRF model) were analysed and compared to Quick Scatterometer (QuikSCAT) satellite surface wind speed observations in order to assess the performance of the former datasets. In [5], the offshore wind power potential based on ocean surface wind fields obtained from the National Oceanic and Atmospheric Administration (NOAA) Blended Sea Winds was estimated. In a more recent study [6] focusing on the development of large-scale offshore wind projects in the Mediterranean Sea, 35-year wind speed data from Modern-Era Retrospective analysis for Research and Applications (MERRA) simulations were used while in [7], the Mediterranean wind climate was described based on the ERA-Interim (ECMWF) dataset covering a rather short period (2015–2018). In the same area, estimates for offshore wind power potential have been also provided in [8], by means of WRF simulation for the period 1979–2016 in the context of the simultaneous exploitation of offshore wind and wave. Although the patterns of the spatial distribution of offshore wind speed and wind power potential provided in the above-mentioned works are very similar, the corresponding estimates may differ due to the different wind data sources used. For instance, satellite observations tend to overestimate wind speed [9]. Finally, in [10], the offshore wind power potential over the Mediterranean was examined under current and future climatic conditions.

Regarding offshore solar energy, floating solar systems at sea are very limited at the moment, with the first one being installed in 2014 in the Maldives, while floating installations on freshwater (lakes, reservoirs) had begun since 2007 (California, USA, and Aichi, Japan). One important advantage of offshore solar energy compared to onshore is that the seawater surface contributes to the increase of efficiency of PVs due to the lower operating temperatures [11], rendering the oceans an attractive alternative solution for installing solar energy devices. Specifically, it has been shown that an increase of the operating temperature by 1 °C leads on average to a decrease of 0.45% in terms of conversion efficiency [12]. Other advantages of offshore solar power plants include the lack of available land space, especially in high-density residential areas where the demands for energy supply are higher and areas with intense agricultural production, and the cleanness of offshore solar panels, due to the reduction of dust adhesion, compared to land-based ones; for a review, see [13]. Nevertheless, offshore floating

PVs still face technical constraints attributed to the harsh sea conditions.

Offshore solar energy has been studied as a standalone form of renewable energy resource in the coastal and offshore environment during the last few years. In the onshore and offshore part of the Mediterranean, offshore solar energy has been estimated in [14] by using numerical results of global horizontal irradiance (between 1991 and 1993) based on the daily mean values from the International Satellite Cloud Climatology Project (ISCCP). The obtained results showed that the solar irradiance in the Mediterranean Sea is higher than the majority of the onshore European part of the relevant countries. A magnitude of solar radiation 5.3 kWh/m$^2$/day was estimated in [15] for the islands of Malta (Mediterranean Sea), showing through an economic analysis that the adoption of offshore PVs could lower electricity costs. Offshore solar energy has been assessed for other marine areas of the world, e.g., in [16], a preliminary estimation of offshore solar energy potential that is available in the exclusive economic zone of India has been presented by using the ERA-Interim solar radiation data. Risk assessment studies on power production from offshore PVs have been provided in [17] for various projects in China based on fuzzy theories, and in [18] for Zhejiang Province (eastern China) under a multi-criteria group decision-making framework.

1.2   Co-exploitation of marine renewable energies

A common challenge for marine energy installations is to reduce the higher construction and maintenance costs compared to the onshore installations. Marine energy installations operate in harsh environmental conditions and thus require materials resistant to erosion and fatigue, while maintenance issues are greatly dependent on the prevailing metocean conditions and distance from shore.

To date, examples for the joint exploitation of the marine resources refer mainly to offshore wind combined with wave energy converters. Specifically, in [8], the authors assessed the combined offshore wind and wave energy in the Mediterranean Sea by introducing an exploitability index. In [19], the variability of the European offshore wind and wave energy resources has been studied and the most promising areas for their combined exploitation have been identified while in [20], the joint exploitation of the resources has been studied for the Italian seas based on a marine spatial planning framework. In [21] and [22], an approach for the identification of the most prominent sites for co-located wave and wind farms has been presented by introducing a co-location feasibility index, and applied this index for a case study off the Danish coast.

In [23] and [24], the authors highlighted that collocating offshore wind and wave might result in reduction of the investment and operational and maintenance costs, or smoother power output; thus the ocean space is more effectively exploited by increasing renewable energy yield and decreasing the overall project cost. From a technical point of view, the combination of two (or more) marine renewable sources may reduce also the impacts of the intermittent nature of these sources; thus the overall marine energy yield in a site is significantly increased delivering a more consistent form of power. From the environmental aspect, the sharing of the ocean space and infrastructure when combining two or more energy resources may have less environmental impacts overall when compared to stand alone installations [23]. Moreover, in [24], the authors showed that the efficiency of two different sources (wind and wave in their study) is increased when there is weak correlation.

Some recent advancements in solar energy sector brought also offshore PVs into play (see e.g., [25]). The combination of PVs with wave energy devices [26] and offshore wind turbines [27] forms a new field of research. In [28], the authors evaluated solar energy and ocean thermal energy conversion in offshore areas of the Caspian Sea, using data from the ERA-Interim database. A similar study was also conducted in [29] for assessing power production from solar and wave energy technologies solely and their combined use as well. New concepts related to the co-exploitation of offshore wind and solar energy by utilizing the same marine space for offshore PV farms and offshore wind farms (OWFs) have very recently appeared ([1]).

1.3    Complementarity of marine renewable energies

Complementarity is one of the most important aspects in the feasibility studies for hybrid or collocated energy installations development. According to [30], "complementarity should be understood as the capability of [different renewable energy sources] working in a complementary way". For instance, there is complementarity in the power production from the combination of wind and solar power plants since PVs produce more energy during the summer months while wind farms reach their maximum electricity production mainly during winter.

Three distinct types of complementarity can be identified according to the particular spatial and temporal framework: i) spatial complementarity; ii) temporal complementarity, and; iii) spatio-temporal complementarity. The quantification of energetic complementarity can be based on a variety of indices, metrics, and alternative approaches. The most widely used family of indices includes various correlation coefficients (CC); the most well-known is the Pearson CC. The Kendall's $\tau$ (tau) and Spearman's rho CCs have been also used in the relevant literature, with the significant advantage of not requiring normality, linearity, and homoscedasticity properties for the examined variables. Specifically, in [31], where the complementarity of wind-solar energy was studied in China, the corresponding joint cumulative distribution function (cdf) has been constructed by using copula theory and Kendall's $\tau$ was used as a measure of the complementarity. For assessing the dependence between three and two different renewable energy sources (wind, sun, and hydro for the first case and wind and hydro for the second), the same coefficient has been used for the description of the non-linear character of the correlation between the outputs of these sources in [32] and [33], respectively. In [34], Spearman's rho has been used along with the Pearson CC for the characterization of the strength of complementarity between hydro and wind energy in Brazil. According to [34], correlation values in the interval (-0.3, 0.0) suggest weak complementarity, (-0.6, -0.3) moderate complementarity, (-0.9, -0.6) strong complementarity, and (-1.0, -0.9) very strong complementarity. Similarly, the corresponding intervals with positive correlation values characterize the strength of synergy. On the other hand, it should be emphasized that since Kendall's $\tau$ and Spearman's rho CCs are non-parametric and refer to the ranks of the examined data, instead of the observed values, they are less powerful and do not use the fine information that is available in a sample (note that the Pearson correlation uses both the mean and the deviation from the mean). A series of time-complementarity indices has been proposed by [35], namely: i) partial time-complementarity index; ii) partial energy-complementarity index, and; iii) partial amplitude-complementarity index. See also the review in [30].

---

([1]) See for example: https://www.pv-magazine.com/2019/12/12/offshore-pv-system-goes-online-in-north-sea/, https://www.greentechmedia.com/articles/read/floating-solar-gears-up-for-the-high-seas, https://www.offshore-mag.com/renewable-energy/article/14182294/offshore-ravenna-wind-farm-combined-with-floating-solar-technology, https://www.offshorewind.biz/2020/08/03/hollandse-kust-noord-to-add-floating-solar-panels-in-2025/, https://www.pv-magazine.com/2020/05/26/solar-arrays-on-wave-energy-generators-along-with-wind-turbines/. Date of last access: 9/11/2020.

The co-exploitation and complementarity of onshore wind and solar energy has been investigated for several regions and locations. In [36], the correlation of wind and solar power with the load demand in Sydney airport has been studied. By assessing the joint availability of the examined sources under the same energy demand together with the corresponding correlation, it was concluded that there is a high complementarity of the resources for the examined location. In [37], a large-scale solar and wind power analysis for Sweden is performed under a scenario of a combined solar and wind power amounting to a total mean annual energy production of 10 TWh. It was found that the combination of solar and wind power generation provides a more balanced distributed output than the individual sources alone, since the resources were negatively correlated on all time scales. In [38], an annealing algorithm has been developed and applied for the Iberian Peninsula in order to identify the optimal locations for wind and solar farms development, in a way that keeps inter-annual and seasonal temporal variations as low as possible and achieves a predefined energy production. In [39], a Monte Carlo simulation has been performed in order to examine wind-sun complementarity in Italy.

In [40] and [41], the complementarity of the resources has been assessed for Europe at different time scales. Specifically, in [40], the authors analysed high-resolution wind-solar data covering three years, while in [41], a new method has been introduced in order to optimise the spatial distribution of wind and solar capacity on onshore and offshore areas of Europe, by minimising the residual demand and assuming a number of different scenarios (e.g., technologies for offshore wind, future electricity demand) for both resources. In [42], the complementarity is studied (at the monthly scale) by utilizing high-resolution maps that are generated based on measurements from weather stations in Mexico. An extended and in-depth complementarity-synergy analysis for Australia has been performed in [43], where several complementarity issues have been discussed for the different examined time scales. Similar extended analyses have been performed for China [44] and Germany [45]. In [44], the complementarity has been studied for both onshore and offshore locations. It was found, among others, that the resources are always negatively correlated at the hourly, daily, and monthly time scales. In [45], the authors studied analytically the spatiotemporal complementarity at various time scales. Since the daily complementarity was generally low, it was concluded that the selection of potential sites characterized by a high degree of complementarity should be made very carefully. In [46], the onshore wind-solar energy synergy has been studied for West Africa. To cope with the limitations of correlation-based approaches and to effectively quantify the synergy, a stability coefficient has been introduced and applied. According to this coefficient, it was concluded that hybrid exploitation of solar/wind power could be beneficial for a large area of West Africa. In [27], a floating hybrid system combining offshore wind and solar resources in Asturias (northern Spain) was studied. A power-smoothing index was proposed to quantify the smoothing in the power output variability from the combination of offshore wind turbines and PV systems, concluding that the power output variability can be reduced up to 68% when compared to wind farms only. A review on the complementarity between grid-connected solar and wind power farms was presented in [47].

In order to generate and provide electricity at large quantities in the future from the co-exploitation of offshore wind and sun, the thorough assessment of the corresponding resource potentials and the involved variabilities is necessary. According to the particular geographical location and considered time scale, wind power density (WP) and solar irradiance are characterized by their spatial and temporal variability [48, 49]. Furthermore, the ideal location and layout of a hybrid marine energy farm along with the identification of the optimal share of offshore wind and solar power that guarantees a high and almost constant (in time) efficiency, are key features in the design phase.

Although many studies focus on the variability and complementarity of onshore wind and solar irradiance, to the best of the authors' knowledge, there is no published study on the joint variability and modelling of offshore wind and solar irradiance at large spatial scales for any marine area around the world. The main aim of this study is to analyse offshore wind and solar resource in the Mediterranean Sea with the ERA5 dataset that is of relatively high spatial and very high temporal resolution. In this context, potentially favourable locations for the co-exploitation of offshore wind and solar energy are identified, taking into account individual and joint variability, and complementarity features. This assessment takes place at the hourly, seasonal and annual scales.

The structure of the present work is the following: In Section 2, the theoretical background and the mathematical tools used herewith are described in detail. In Section 3, the ERA5 reanalysis dataset for assessing offshore wind and solar resource is presented while in Section 4, the univariate assessment of WP and solar power (SP), and the relevant variabilities in the seasonal and annual scales is provided. In the same section, the joint coefficient of variation is implemented here for the first time. Section 5 deals with a preliminary assessment of some important aspects of WP and SP, such as synergy, complementarity, non-availability, and intermittency that are mathematically described in Section 2.3. The relevant analysis is based on an event-based approach. Association measures are also analysed and compared, including the robust CC $r_{MED}$ along with more traditional tools for complementarity analysis, such as the Kendall's $\tau$ and the Pearson's $r$ CCs. Moreover, the monthly energy output is estimated for selected locations to assess the possibility of deploying hybrid plants in the Mediterranean basin. The last section addresses the concluding remarks of this analysis and suggestions for further research.

## 2. THEORETICAL BACKGROUND

In the univariate and bivariate assessment of wind and solar energy, the following aspects are of particular importance:

**Time scales.** The assessment of the resource at different time scales reveals different patterns and provides important information for the examined resource. In this work, the following time scales are considered: i) the finest time scale of the examined variable, which corresponds to the recording interval. In ERA5 dataset (see section 3), the finest available time scale corresponds to one hour, and the examined variables refer to instantaneous (i.e., hourly) power values. ii) The seasonal time scale, where the examined variable is the mean value of the hourly observations for each season of the time series, referred to as "seasonal mean value (for the particular year)". The sample size of these values corresponds to the total number of seasons of the initial time series. iii) The annual time scale, where the examined variable is the mean value of the hourly observations for each year of the time series. This variable is subsequently referred to as "annual mean value (for the particular year)".

Two additional time scales that may be used are the following: i) the daily time scale, where the examined variable is the mean value of the hourly observations for each day of the time series (referred to as "daily mean value"), and; ii) the monthly time scale, where the examined variable is the mean value of the hourly observations for each month of the time series (referred to as "monthly mean value").

It is evident that according to the working time scale, a "new" variable (produced through averaging of subseries of the initial time series) arises each time. For example, in the daily time scale, the new variable is the daily mean value (or simply "daily value"). These new variables are characterized by their corresponding statistical (sample) parameters, i.e., mean value, standard deviation, coefficient of variation, etc. These parameters will be subsequently referred to as "mean daily value", "standard deviation of daily values", "coefficient of variation of daily values", etc. See also the relevant discussion in [50].

**The variability of the resource(s).** The individual variability of a resource is generally and traditionally expressed in the spatial domain through static measures, such as the mean annual variability ($MAV$) and inter-annual variability ($IAV$) and, more generally, via an appropriately defined coefficient of variation ($CV$). Let it be noted though, that in resource analysis for a particular location, the available data series could be examined through an analytical time series model or spectral analysis and provide much finer and detailed information on the variability of the series itself. On the other hand, for gridded data in extended spatial scales (such as the Mediterranean Sea), the variability measures based on $CV$ are directly attainable for any examined time scale, while they provide a unique, yet static, baseline for comparison purposes in the spatial domain. See also [51], [52], and [53] and references cited therein. In addition, a joint variability measure, namely the joint coefficient of variation ($JCV$) for WP and SP is applied here for the first time. See section 2.1.

**Complementarity and synergy.** CC is extensively used in the literature for the characterization of the strength of complementarity between two renewable energy sources. In principle, CC is used to quantify the association/correlation between two random variables. CC takes values in [−1,1], quantifies the strength of association between the random variables, and reveals the direction of this relationship. Positive values denote a similar behaviour between the two random variables; for instance, if the one random variable is increasing (decreasing), then the other is also increasing (decreasing). Negative values (called also anti-correlation) suggest an opposite behaviour, while the association between the random variables becomes weaker as the values of CC tend to 0. Consequently, in the assessment of hybrid renewable energy sources, negative values of CC suggest complementarity while positive values indicate synergy between the sources. There are various forms of CC that are presented in section 2.2. Moreover, in [45], use of the t-copulas copulas parameters has been made in order to assess inter-regional wind-solar complementarity (on a daily and mean annual basis).

## 2.1 Variability measures

The most common variability measures for a random variable $X$ are the standard deviation $\sigma_X$ and the coefficient of variation $CV_X$, i.e., the ratio of standard deviation $\sigma_X$ to the mean value $\mu_X$. The use of $CV$ in resource analysis using long-term time series of the examined parameters has been used in a great extent in the relevant literature; see e.g., the recent review [30].

However, $CV$ is sensitive to the presence of outliers or for distributions with long tails (such as WP) leading to overestimated values of $CV$. Note that neither $\sigma_X$ nor $\mu_X$ are robust or resistant, [54]; therefore, in this work, a robust coefficient of variation $RCV$ based on the median absolute deviation ($MAD$) is also implemented. $RCV$ is defined as follows:

$$RCV_X = \frac{MAD}{\hat{x}_{0.5}}, i = 1, 2, \cdots, N, \qquad (1)$$

where $MAD = \text{med}|x_i - \hat{x}_{0.5}|$, med(·) denotes the median, $\hat{x}_{0.5}$ is the 0.5 quantile and $x_i, i = 1, 2, \cdots, N$, are the values of the examined variable, [30], [55]. Another widely used robust coefficient of variation is based on the interquartile range (IQR); however, $RCV_X$ is preferred since it has smaller variability, [56]. The use of $RCV$ in wind energy assessment studies is strongly suggested in [54], after a thorough evaluation of 27 different methodologies on the same 37-year monthly wind speed and energy production time series. In the same work, it is also noted that "Despite the importance of long-term variability, the wind-energy industry lacks a systematic method to quantify this uncertainty." See also [57], where $RCV$ has been used for the variability assessment of the wind power density for the USA.

Two traditionally used temporal variability measures in resource assessment are the mean annual and inter-annual variability ($MAV$ and $IAV$, respectively) that are actually defined in a similar way as $CV$. $MAV$ is the ratio of the mean value of the ratio of the annual standard deviation to the annual mean value of the variable $X$ for a specific time period of $J$ years, while $IAV$ is defined as the ratio of the standard deviation $s_{m_X(j)}$ of the annual mean values of the variable $X$, $m_X(j)$, $j = 1, 2, \ldots, J$, to the overall mean value $\widetilde{m}_X$; see also [50].

Two additional variability indexes are the monthly and seasonal variability ($MV$ and $SV$, respectively), [58]. These indexes present the variations of $X$, which occur from one month to another ($MV$) and from one season to another ($SV$). $MV$ and $SV$ are defined as follows:

$$MV = \frac{X_{M_{max}} - X_{M_{min}}}{\bar{X}_{an}}, SV = \frac{X_{S_{max}} - X_{S_{min}}}{\bar{X}_{an}}, \qquad (2)$$

where $X_{M_{max}}$ ($X_{S_{max}}$) denotes the maximum mean monthly (seasonal) value of $X$, $X_{M_{min}}$ ($X_{S_{min}}$) denotes the minimum mean monthly (seasonal) value of $X$, and $\bar{X}_{an}$ denotes the mean annual value of $X$.

In the case of a multivariate random variable $X = (X_1, X_2, \cdots, X_m)$, various, rather unknown, definitions have been proposed for the multivariate coefficient of variation $JCV_X = JCV_{X_1 X_2 \cdots X_m}$. See also the review in [59]. In order to define $JCV_X$, let $x_{1,i}, x_{2,i}, \cdots, x_{m,i}$, $i = 1,2,\cdots,n$, denote the corresponding sample values, $\bar{x} = (\bar{x}_1, \bar{x}_2, \cdots, \bar{x}_m)$ the vector of sample mean values, and $S$ the sample covariance matrix. According to [60], the sample multivariate coefficient of variation of $X$ is estimated as follows:

$$JCV_{X(VN)} = \sqrt{\frac{1}{\bar{x}^T S^{-1} \bar{x}}}, \qquad (3)$$

which for two random variables is simplified as follows:

$$JCV_{XY(VN)} = \sqrt{\frac{s_X^2 s_Y^2 - s_{XY}^2}{\bar{x}^2 s_Y^2 - 2\bar{x}\bar{y}s_{XY} + \bar{y}^2 s_X^2}}. \qquad (4)$$

$JCV_{X(VN)}$ provided by equation (3) is invariant under change of scale and reduces to the univariate $CV$ when $m = 1$. Other definitions of $JCV_X$ can be found in [59]. Following the suggestions of Aerts et al. [59], in this work relation (4) is applied for the estimation of $JCV$ of WP and SP.

2.2 Correlation metrics

Among the different types of CCs, the Pearson $r$ is the most widely used. The main assumptions underlying its implementation are the following: i) the random variables are normally distributed; ii) there are no outliers present in the sample, and; iii) linearity and homoscedasticity should hold true. Therefore, Pearson correlation is not appropriate for the case of a nonlinear relationship between the variables. In [44], the use of $r$ is not suggested for complementarity analysis of wind and solar resources, since these resources are not necessarily associated in a linear way. See also [61], where it is emphasized that the use of $r$ is "catastrophically bad under contamination", i.e., for contaminated normal data. In [62], a detailed evaluation as regards the performance of the Pearson CC compared to robust CCs is made for both bivariate normal and contaminated normal distributions. A main conclusion of that study was that "the behavior of the sample correlation coefficient $r$ is disastrous under contamination". However, in resource analysis of complementarity, there is not yet a consensus as regards the use of the "correct" CC.

The sample version of $r$ is provided by:

$$r = \frac{\sum_{j=1}^{N}(X_j - \bar{X})(Y_j - \bar{Y})}{\sqrt{\sum_{j=1}^{N}(X_j - \bar{X})^2}\sqrt{\sum_{j=1}^{N}(Y_j - \bar{Y})^2}}, \qquad (5)$$

where $\bar{X}, \bar{Y}$ are the mean values of $X$ and $Y$, respectively, and $X_j, Y_j$ is the individual $j$−th sample point of $X$ and $Y$, respectively, with $j = 1, ..., N$.

Two well-known non-parametric CCs are the Kendall's $\tau$ and Spearman's rho. As mentioned before, Kendall's $\tau$ and Spearman's rho are less powerful compared to the Pearson CC and neglect the fine information that is included in the data sample.

In order to estimate the sample (empirical) version of Kendall's $\tau$ CC, the values of $X_i$ should be first ordered in increasing order, and then the number of corresponding values of $Y_i$ that satisfy (concordant pairs) or not (discordant pairs) this order should be counted. The sample estimate of Kendall's $\tau$, $\hat{\tau}$, is provided as follows:

$$\hat{\tau} = \frac{2}{n(n-1)}(N_C - N_D), \qquad (6)$$

where $N_C$ and $N_D$ denote the number of concordant and discordant pairs, respectively, in the $(X_i, Y_i)$ sample of observations. When ties exist in the data, a modified version of equation (6) should be used.

Another category of CCs refers to robust estimators. As is noted in [61], robust methods secure the stability of the statistical inference when there are deviations from the assumed distribution model. In the same work, various robust estimators for the CC are reviewed and evaluated against the Pearson CC. For the random variables $X$ and $Y$, the correlation median estimator $r_{CMED}$ is defined as follows:

$$r_{CMED} = \frac{\text{med}^2|u| - \text{med}^2|v|}{\text{med}^2|u| + \text{med}^2|v|}, \qquad (7)$$

where

$$u = \frac{x - \text{med}x}{\text{med}|x - \text{med}x|} + \frac{y - \text{med}y}{\text{med}|y - \text{med}y|}, v = \frac{x - \text{med}x}{\text{med}|x - \text{med}x|} - \frac{y - \text{med}y}{\text{med}|y - \text{med}y|}, \qquad (8)$$

and $\text{med}(\cdot)$ denotes the median operator. See also [63].

It is evident that the use of each CC is based on the underlying assumptions that should be checked carefully before any realistic application. Clearly, the use of a unique CC could be safely decided when a potential area for hybrid development should be assessed in detail. In this case, the presence or lack of normality, linearity, and homoscedasticity in the bivariate data set would dictate the specific CC to be used. In section 5.1, results referring to Pearson's $r$,

Kendall's $\tau$ and $r_{CMED}$ are provided, compared, and discussed for the annual, seasonal, and hourly time scales.

2.3   Event-based complementarity aspects

Correlation coefficient provides an indication of the degree of complementarity between WP and SP. In [43], alternative and more representative measures that rely on predefined, realistic quantities relevant with the feasibility of wind and solar energy development in the spatial scale are provided. Specifically, in order to assess the availability, complementarity, synergy, and persistence of offshore wind and solar energy, a lower threshold for WP ($WP_L$) and SP ($SP_L$), are firstly introduced. $WP_L$ and $SP_L$ are treated as cut-off values, below of which the efficient exploitation of the corresponding resources becomes, in principle, uncertain.

Let $WP_{AN}$ and $SP_{AN}$ denote the mean annual wind and solar irradiance power density, respectively. The single events that can be considered in complementarity analysis are the following:

$$E_W = [WP > WP_L], E_W^C = [WP \leq WP_L], E_{W,AN} = [WP_{AN} > WP_L], \qquad (9)$$

$$E_S = [SP > SP_L], E_S^C = [SP \leq SP_L], E_{S,AN} = [SP_{AN} > SP_L], \qquad (10)$$

where $E^C$ denotes the complementary event of $E$. In this setting, the following indexes can be defined:

1. Wind to solar power complementarity, $WCS$:

$$WCS = \Pr\left[E_W \cap E_S^C\right]. \qquad (11)$$

2. Solar to wind power complementarity, $SCW$:

$$SCW = \Pr[E_W^C \cap E_S]. \tag{12}$$

Large values of $WCS$ ($SCW$) suggest that wind (solar) strongly complements solar (wind) and vice versa. A value of $WCS$ equal to unity suggests full complementarity between the two sources, while $SCW$ cannot approach unity due to the night-time effects, where SP is zero. A low value of $SCW$ denotes very poor solar to wind complementarity.

3. Joint non-availability of wind and solar power, $UWS$:

$$UWS = \Pr[E_W^C \cap E_S^C]. \tag{13}$$

Large values of $UWS$ suggest that wind and solar are jointly unavailable in a great temporal extent. A value of $UWS$ close to 1 indicates that the specific area is out of consideration regarding the joint development of offshore wind and solar.

4. Synergy of wind and solar power, $SWS$:

$$SWS = \Pr[E_W \oplus E_S], \tag{14}$$

where $\oplus$ denotes the XOR (eXclusive OR) operator. Areas characterized by $SWS$ close to 1 are highly synergetic while values close to zero suggest that the synergy between offshore wind and solar is very poor.

The persistence of events during which the instantaneous (hourly) values of WP and SP are below the threshold levels are also of importance. The duration statistics of these events can be used as indicators of the degree of intermittency of the two energy sources. In addition, to exclude areas that are, in principle, of no interest for hybrid offshore wind and solar development, the above events are conditioned with respect to the mean annual values of WP and SP, respectively. The persistence of WP and SP is then defined as follows:

$$D_{NW} = \text{Duration of the event } [E_W^C|E_{W,AN}], \tag{15}$$

and

$$D_{NS} = \text{Duration of the event } [E_S^C|E_{S,AN}]. \tag{16}$$

$D_{NW}$ and $D_{NS}$ are random variables representing the number of consecutive hours where the events in the right-hand side of relations (15) and (16) are true. Based on the above, the compound event $D_{NW \cap NS}$ can be also defined as follows:

$$D_{NW \cap NS} = \text{Duration of the event } \left[[E_W^C|E_{W,AN}] \cap [E_S^C|E_{S,AN}]\right]. \tag{17}$$

Large values of $D_{NW \cap NS}$ at a location, suggest that although the particular location satisfies the basic resource prerequisites for WP and SP, it should be carefully considered for hybrid development of offshore wind and solar farms since the degree of the joint intermittency is, in this case, high.

3. **ANALYSED DATA**

In this study, the recent ERA5 reanalysis dataset, produced by ECMWF [64] and released in 2019, is used. ERA5 is the fifth generation atmospheric reanalyses of the global climate and can be freely accessed from the Copernicus Climate Data Store (https://cds.climate.copernicus.eu/, [65]) through either the web interface or a tailored Python script via CDS API in netcdf format. The most considerable improvements, with respect to ERA-Interim dataset, refer to the finer spatial and temporal resolution (79 km vs 31 km for the horizontal dimension, 60 levels vs 137 levels for the vertical dimension and 6 h vs 1 h, respectively), the upgrade of the assimilation system and the radiative transfer model, the larger period covered, and the better representation of physical processes (e.g., tropical cyclones, better balance of precipitation and evaporation). Apart from the technical upgrades, another major reason for using this specific dataset is the fact that solar irradiance and sea surface temperature are obtained from the same physical model and hence, physical and spatial consistency is established at every time step of the available time series.

The sufficient spatial and temporal resolution of ERA5 data renders this dataset quite appropriate in preliminary wind energy studies. For instance, it has been used for offshore wind power assessment studies in the Caspian Sea [66]. Sea surface wind speed from ERA5 was compared and evaluated against three additional datasets obtained from Advanced Scatterometer (ASCAT), ERA-Interim and measurements from an offshore platform. ERA5 found to underestimate wind speeds below 2 m/s. In [67], the offshore wind technical power potential was estimated for southeast and south coastal regions of Brazil. These authors evaluated three different atmospheric reanalysis databases (CFSv2, ERA5, and MERRA2) by using *in situ* measured wind data from buoys. It was concluded that ERA5 performed better for all examined locations. In [68], ERA5 has been used for the estimation of the global wind power potential taking into account the seasonal variability of air density, and in [69], the performance of ERA5 data in wind power modelling has been evaluated against the MERRA-2 dataset using aggregated wind energy generation at a country (for five countries) and individual wind turbine (for 1051 wind turbines in Sweden) level. By using CC, mean absolute and root mean square error as evaluation metrics, it was found that ERA5 provided more accurate estimates than MERRA-2 for all considered metrics. In [70], ERA5 wind speeds at different heights above ground have been evaluated by using measurements obtained from lidars. It was found that the relative deviation of the ERA5 data from the measurements was below 20% at all examined heights. Other uses of ERA5 data in wind energy assessment can be found, for example, in [71] (where the effects of the gravity waves induced from an offshore wind farm in the annual energy production are studied) and [72] (where an airborne wind power density analysis is made for different ceiling heights greater than the current standard hub heights).

ERA5 solar irradiance estimates have been evaluated in [73], by using satellite products and stations all over the world. The results of that analysis showed that ERA5 dataset achieves a reduction in the bias compared to ERA-Interim and MERRA-2, mitigating the quality gap with satellite-derived data. Nevertheless, the current resolution for areas with highly varying solar irradiance (e.g., coastal areas) is still insufficient. Furthermore, the same dataset has been implemented in [74], where the surface diffuse solar radiation was evaluated over East Asia using ground measurements. The obtained results indicated that ERA5 underestimated surface diffuse solar radiation while its performance on a monthly scale was better compared to the daily and hourly scale. In [75], this dataset was validated, among seven other reanalysis and satellite-derived data, against multiple ground stations around the world for a 27-year period. It was clearly stated that radiation data from ERA5 is superior than MERRA-2 with reference to the examined error metrics. Although some improvements are still needed, such as in the modelling of cloud properties [76] and aerosols, the ERA5 gridded solar irradiance data perform overall satisfactorily among other reanalysis products, e.g., for Australia [77] and Norway [78]. Let it be also noted that, in the present study, the Mediterranean Sea presents more often clear-sky conditions than other European areas, resulting to less variability [79].

The variables analyzed in this study include the wind speed (in m/s) at 100 m above sea level, and the surface solar radiation downwards (in $J/m^2$). The latter variable expresses the amount of solar energy (both surface direct and diffuse) that reaches a horizontal plane at the Earth's surface. Solar irradiance was converted to W/m² for homogeneity reasons in terms of units. The analyzed time series were extracted directly from the ERA5 dataset for the entire available period 1979–2019 (i.e., 41 years). This length is very appropriate for the description of statistical patterns that are present in the hourly, daily, monthly, seasonal, annual, and inter-annual time scales.

## 4. OFFSHORE WIND AND SOLAR RESOURCE AND VARIABILITY

In this section, the univariate assessment of the offshore wind and solar resource at the Mediterranean basin level is presented. Two main aspects are analysed: the available resource in various time scales and the corresponding variability.

The overall mean wind (solar) power density $\overline{WP}$ ($\overline{SP}$, in W/m²) is directly estimated by utilizing the time series of hourly wind power density (solar irradiance) at the model grid points. For a particular grid point, $\overline{WP}$ is estimated by the following equation:

$$\overline{WP} = \frac{1}{N} \sum_{i=1}^{N} WP_i, \qquad (18)$$

where $WP_i = \frac{1}{2} \rho u_i^3$, $i = 1, 2, \cdots, N$, and $u_i$ is the wind speed at 100 m above sea level at the examined time scale, $\rho$ is the air density, considered to be equal to 1.2258 kg/m³, and $N$ is the corresponding sample size of the time series.

In Figure 1, the sea areas that are mentioned regularly in the remaining of the text are depicted.

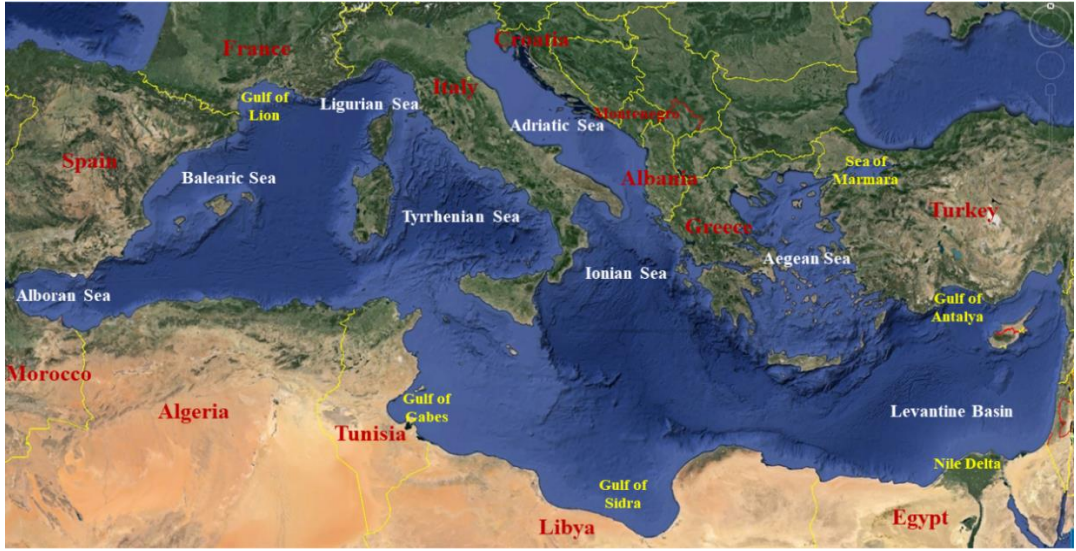

Figure 1. Sea areas in the Mediterranean Sea mentioned in the text.

4.1  Offshore wind resource

In Figure 2 (upper panel), the mean annual WP for the period 1979–2019 is presented. The largest values of WP are depicted in the Gulf of Lion (~1,130 W/m$^2$) and the central and southeast Aegean Sea (around 760 W/m$^2$ and 720 W/m$^2$, respectively). Large values of WP are also spotted in the Alboran Sea (highest value ~745 W/m$^2$ at 36.25N, 3.00W).

In the same figure, the mean seasonal WP is also presented for each season. The most energetic season is winter with the highest value (1,540 W/m$^2$) located in the Gulf of Lion. The extended offshore area between Mallorca and Sardinia Isl., and the northern Aegean Sea are also characterized by high values (around 1,000 W/m$^2$). Despite the fact that summer exhibits the lowest values of WP overall for the examined basin, the etesian winds blowing in the Aegean Sea during this period of the year result in higher WP values (over 870 W/m$^2$) in that area compared to the rest Mediterranean Sea; in the Gulf of Lion, the corresponding values are between 700 W/m$^2$ and 800 W/m$^2$. The spatial distribution of WP during spring and autumn is almost alike; for both seasons, the Gulf of Lion represents the most energetic area reaching values around 1,150 W/m$^2$ and 1,060 W/m$^2$, respectively.

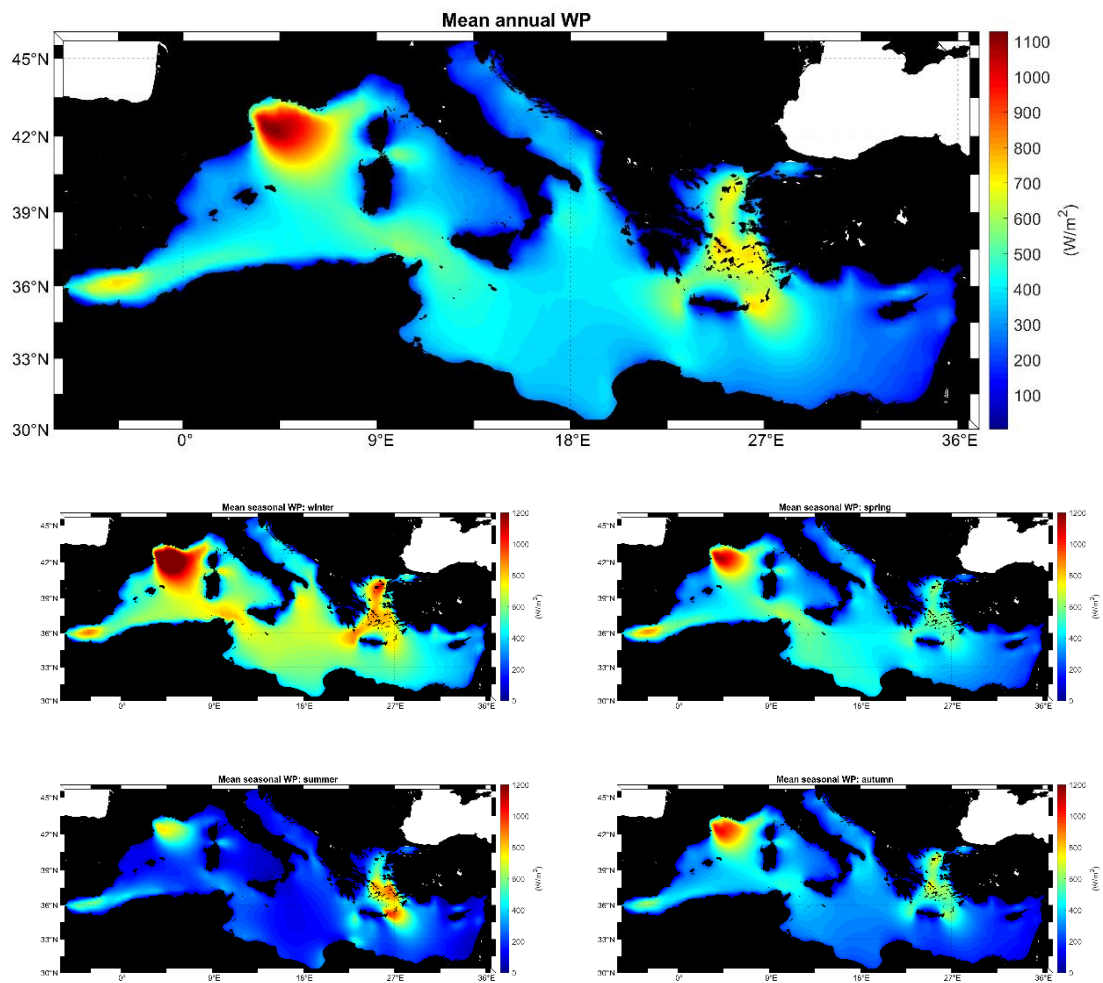

Figure 2. Mean annual (upper panel), and seasonal wind power density for the Mediterranean Sea. Winter: middle left panel, Spring: middle right panel, Summer: lower left panel, Autumn: lower right panel.

In Figure 3, the $50^{th}$ (i.e., the median value), $75^{th}$, $90^{th}$, and $95^{th}$ percentile values of the hourly values of WP are shown.

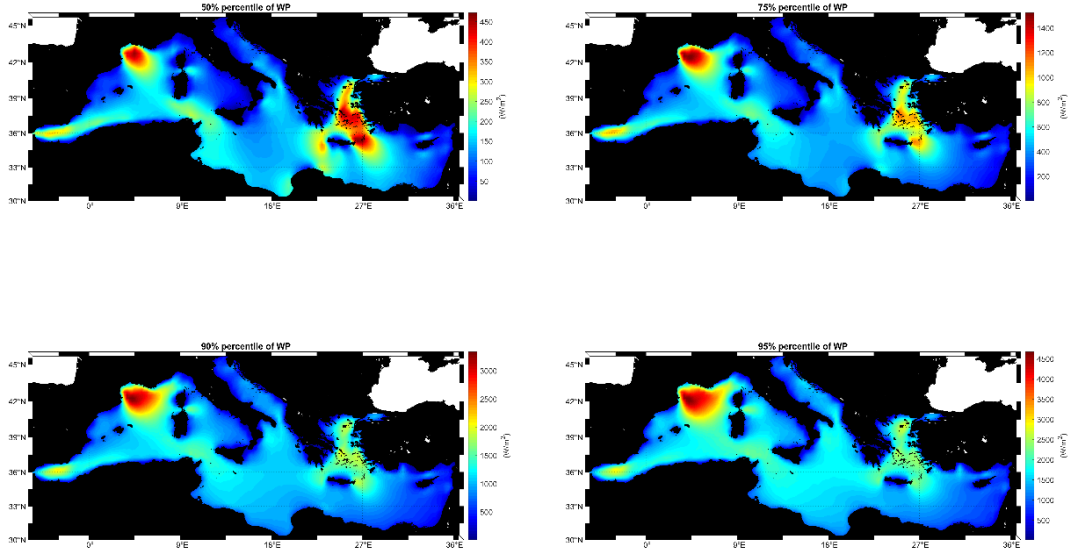

Figure 3. 50th (upper left panel), 75th (upper right panel), 90th (lower left panel), and 95th (lower right panel) percentiles of WP hourly values for the Mediterranean Sea.

Although the spatial patterns are very similar, the differences in the percentile values are very large. This characteristic confirms that the distribution of WP has a long right tail. In addition, there are clear differences in the distribution of the mean annual values (Figure 2) and the median values. As is emphasized in [57], where the detailed characterization of wind power resource in the United States has been made, "*We therefore view the median to be a more robust indicator of central tendency and a more appropriate metric to represent WPD* [wind power density]".

Regarding the variability of offshore wind resource, the spatial distribution of $MAV$ and $IAV$ is displayed in Figure 4 ($^2$). The highest values of $MAV$ are encountered in different limited areas, such as the northern and eastern coasts of Sicily Isl., the north and southwestern coasts of Italy, the southwestern coasts of Corsica Isl., the Gulf of Antalya, the northern coasts of the Aegean Sea and across the eastern Mediterranean coasts. The entire Aegean Sea, the Gulf of Lion, the Alboran Sea and the greatest part of the southern coasts of the Mediterranean are characterized

---

($^2$) Let it be noted that the classical $CV$ of the hourly values of WP is identical to $MAV$ by definition and the $CV$ of the annual mean values is very close to $IAV$.

by relatively low values of *MAV* (of the order of 110–160%). The highest value (320%) is depicted in the Gulf of Antalya while the lowest one (107%) is located in the southern part of the Aegean Sea (35.5N, 26.25E). Other locations with high *MAV* values (over 220%) include the coastal areas of the Ligurian, Tyrrhenian, and Adriatic seas.

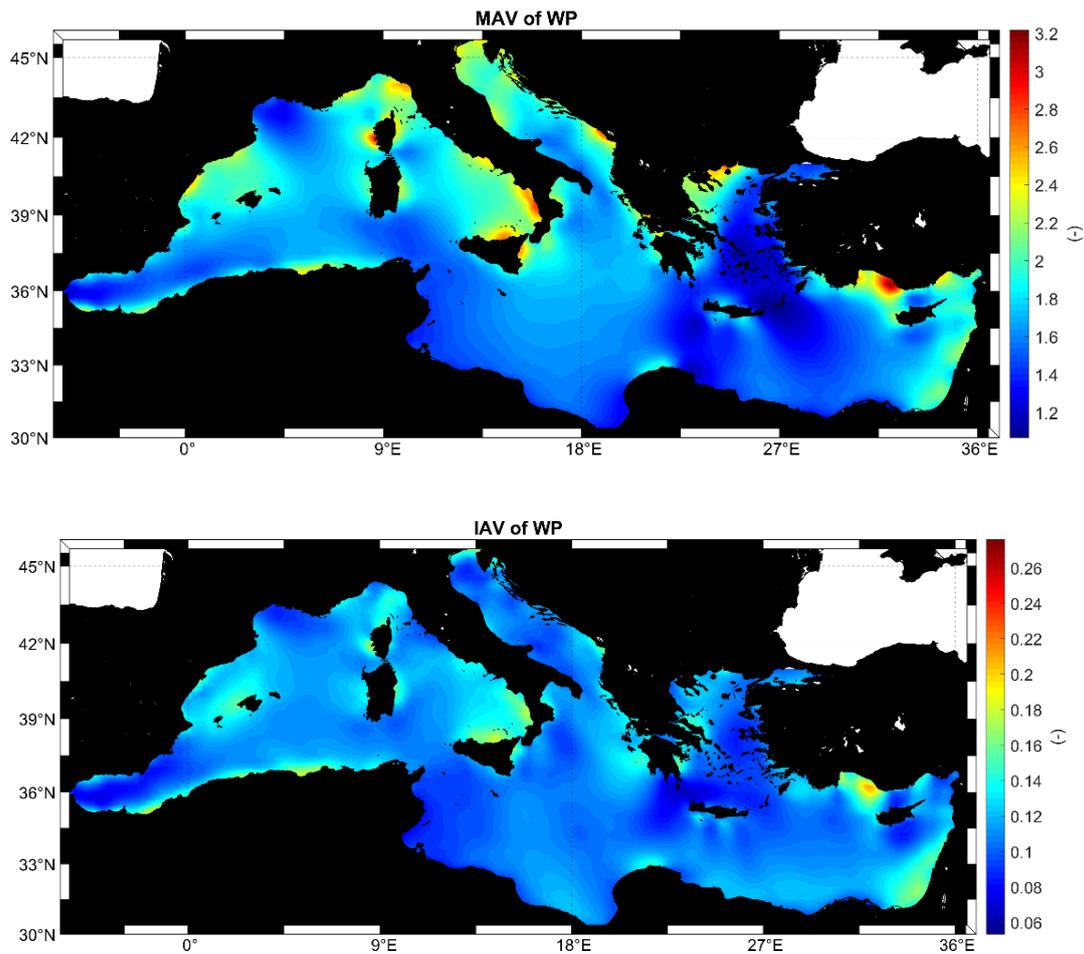

Figure 4. Mean annual (upper panel) and inter-annual (lower panel) variability of wind power density for the Mediterranean Sea.

*IAV* takes generally low values (of the order of 7%–12%). The highest value (21%) is located again in the Gulf of Antalya while southern Tyrrhenian and southeast Levantine seas are also characterized by large values that range from 15% to 18%. On the other hand, the northwestern part of Crete Isl. presents the lowest values of *IAV* (7%), followed by the central Aegean and Alboran seas and the Gulf of Lion.

Overall, between areas with equivalent (and exploitable) WP, the ones with the lower values of $MAV$ and $IAV$ are more preferable for consideration of OWF development. The Gulf of Lion, Aegean, Alboran, and southern Adriatic seas, the Gulf of Sidra, and the strait between Sicily Isl. and Africa, are, in principle, very promising areas for OWF development.

The robust coefficient of variation $RCV$ of the annual and seasonal mean values of WP is depicted in Figure 5. As mentioned before, in [54], the use of $RCV$ is highly suggested for evaluation of the variability in wind energy studies. In the annual scale, the Gulf of Antalya is characterized by the highest value (15%), followed by northern Sicily Isl. (12%). The areas where the lowest values are depicted are the Aegean (3%) and Alboran (4%) seas, the Gulf of Gabes (4%), and southern Sardinia Isl. (3%). Regarding the seasonal scale, the Gulf of Antalya is characterized by the highest values for all seasons, followed by the Ligurian Sea. The southern Aegean Sea and some offshore areas in the central Mediterranean Sea exhibit low values of $RCV$ during the examined seasons. In general, $RCV$ takes relatively high values during winter for a great part of the Mediterranean basin while the western part exhibits a rather stable behaviour (with relatively low values) for spring and summer. Low values of $RCV$ suggest potentially appropriate areas for OWF development with respect to the above variability metric, such as the entire Aegean and Alboran seas, and the gulfs of Lion and Sidra. Note that the same areas are also suggested as potential candidate areas for OWF development based on $MAV$ and $IAV$.

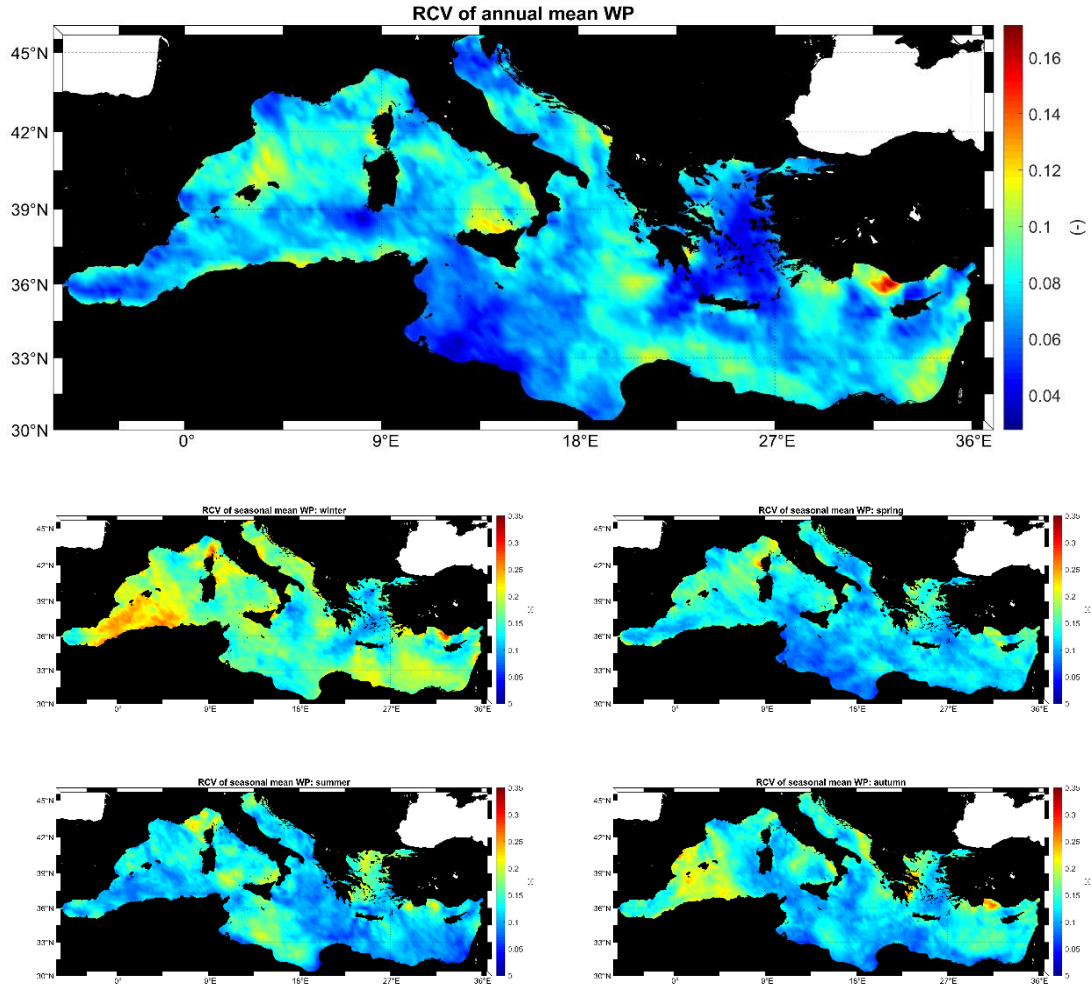

Figure 5. Robust coefficient of variation for the mean annual (upper panel) and mean seasonal wind power density for the Mediterranean Sea. Winter: middle left panel, Spring: middle right panel, Summer: lower left panel, Autumn: lower right panel.

In Figure 6, the spatial distribution of the seasonal variability of WP is presented. In general, monthly variability (not shown here) takes slightly higher values than seasonal variability. The lowest values for *SV* are encountered in the Aegean and Alboran seas and the highest ones in the Gulf of Antalya and the central Mediterranean Sea (mainly Tyrrhenian and Libyan Seas). *SV* takes their lowest values in the southern Aegean Sea (0.32). The values of *MV* and *SV* were also compared to the numerical results shown in Figures 6 and 7 of [8], with wind data obtained from the WRF model between 1979 and 2016. The similarity of the spatial distribution of both indexes is rather strong.

Areas characterized by low values of *MV* and *SV* (such as the central and southern Aegean and Alboran seas, and the gulfs of Lion, Sidra, and Gabes) are appropriate for OWF development with respect to the above variability metrics.

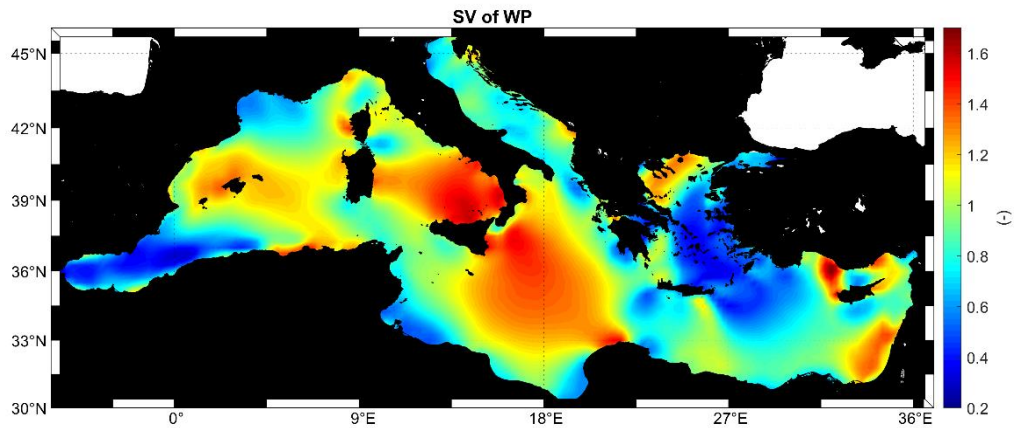

Figure 6. Seasonal variability of wind power density for the Mediterranean Sea.

4.2  Offshore solar resource

The largest mean annual values of SP are located at the southeast part of the Mediterranean Sea, and the peak value is spotted near the coasts of Egypt (~236 W/m$^2$); see Figure 7 (upper panel). The southern Aegean, Levantine, and Alboran seas exhibit large values of SP with corresponding values above 220 W/m$^2$. Lower values of SP are encountered at the northern coasts of Mediterranean Sea, especially in the northern Adriatic Sea (less than 180 W/m$^2$).

In the same figure, the spatial distribution of the mean seasonal SP is presented (middle and lower panels). The most energetic season in terms of SP is evidently summer, with the highest value encountered in the eastern Levantine Sea (320 W/m$^2$). The Libyan, Aegean, and Alboran seas present high values of mean SP, which exceed 310 W/m$^2$. Similar spatial distributions for SP are presented in autumn and spring. During spring, the Levantine Sea is characterized by the largest values, which range between 270 W/m² and 280 W/m² (the maximum value is located off the coasts of Egypt); the corresponding values for autumn are 190 W/m² and 200 W/m² (the maximum value is located off the Nile Delta). The mean SP in winter takes its lowest values (the highest winter value is located at the Gulf of Sidra: ~150 W/m²).

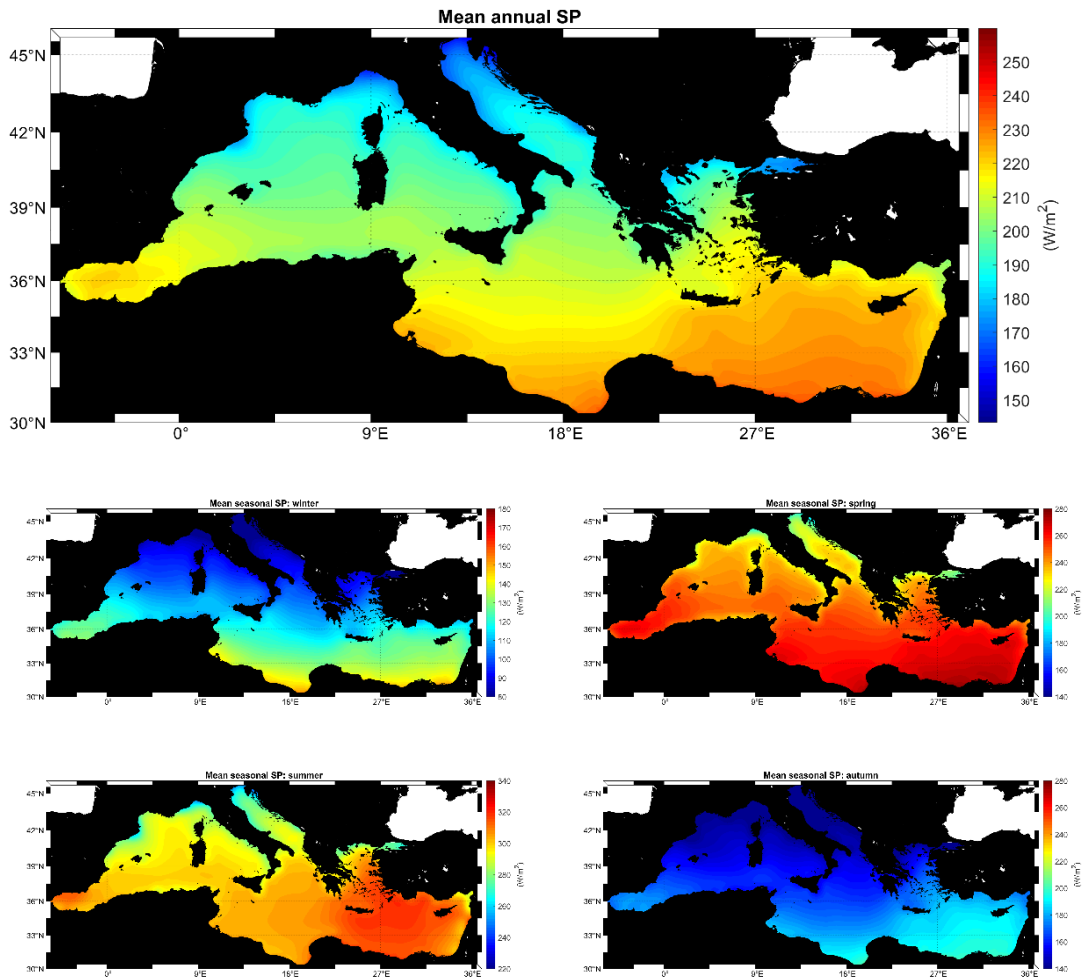

Figure 7. Mean annual (upper panel) and mean seasonal solar irradiance for the Mediterranean Sea. Winter: middle left panel, Spring: middle right panel, Summer: lower left panel, Autumn: lower right panel.

In Figure 8, the 50th (i.e., the median value), 75th, 90th, and 95th percentile values of the hourly values of SP are shown. As was the case with percentile values of WP, in this case also the differences between the different percentile values of SP are evident.

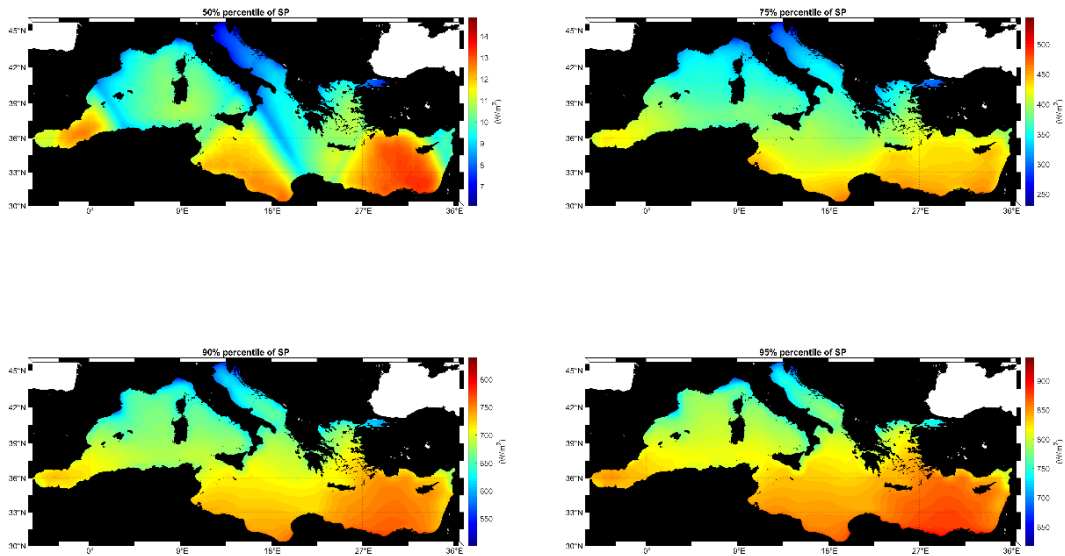

Figure 8. 50$^{th}$ (upper left panel), 75$^{th}$ (upper right panel), 90$^{th}$ (lower left panel), and 95$^{th}$ (lower right panel) percentiles of SP hourly values for the Mediterranean Sea.

The spatial distribution of *MAV* and *IAV* for SP is shown in the upper and lower panel of Figure 9, respectively. *MAV* is clearly lower and exhibits lower spatial variability than the *MAV* for WP. Specifically, the higher values are encountered in the northern coasts of Italy, in the Adriatic Sea, and the western Aegean coasts. The highest value of *MAV* is located in the northern Adriatic Sea (145%) while in the northern Aegean Sea, the values of *MAV* exceed 140%. In general, the values of *MAV* for the southern Mediterranean and Levantine seas are clearly lower than these of the northern Mediterranean. The central and southern Aegean and the Alboran seas, as well as the southern and eastern coasts of the Mediterranean, are characterized by relatively low values of *MAV* (of the order of 133%–137%). The spatial distribution of *IAV* is similar to *MAV*. Relatively large values (around 3%) are encountered in the northern Adriatic Sea, the Gulf of Genoa, and the coastal area of Barcelona. The Balearic, northern Aegean, and eastern Tyrrhenian seas are also characterized by relatively large values of *IAV*. Similar to *MAV*, *IAV* takes low values in the entire Levantine Sea.

Based on the above results, along with the corresponding mean annual values of SP, the southern part of the Mediterranean Sea is clearly more favorable for the development of floating solar panels farms.

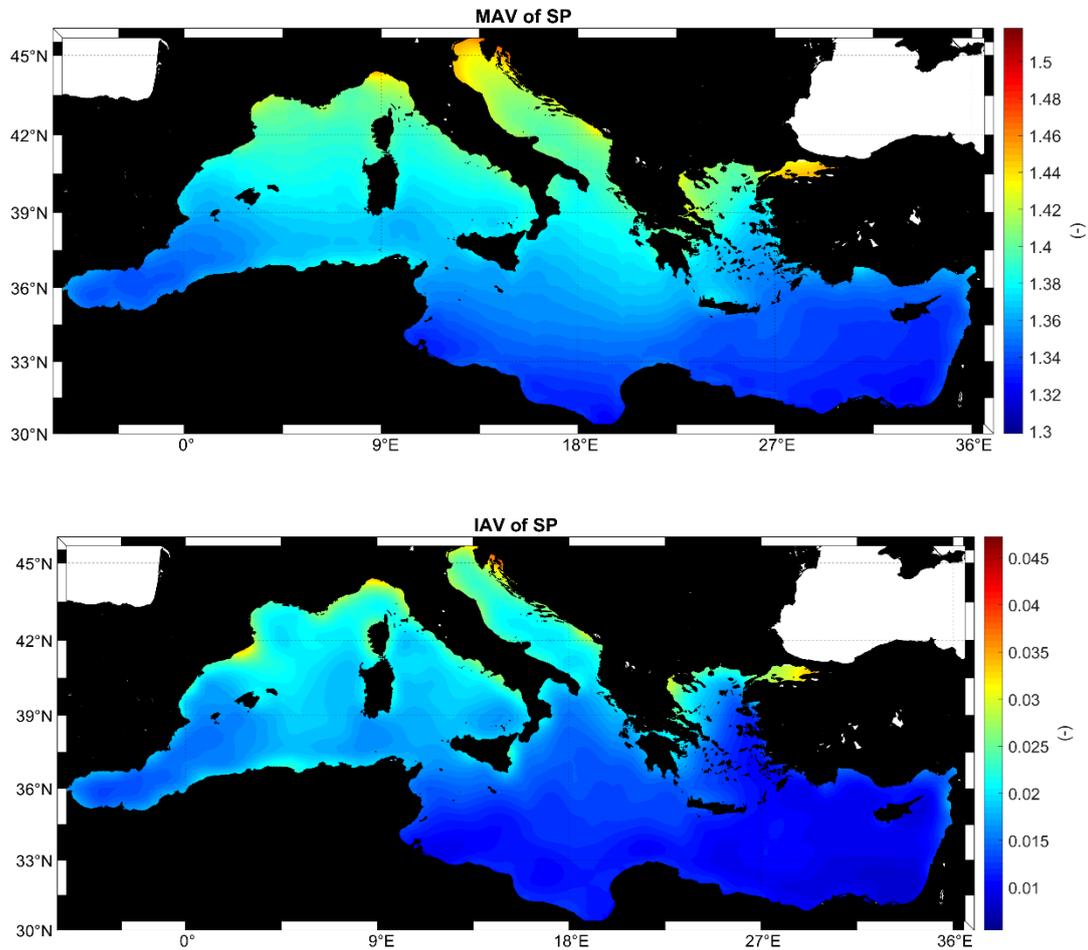

Figure 9. Mean annual (upper panel) and inter-annual (lower panel) variability of solar irradiance for the Mediterranean Sea.

The $RCV$ of the annual and seasonal mean values of SP is depicted in the upper and middle/lower panels of Figure 10, respectively. The highest values of $RCV$ are encountered across the northern coasts of the basin (especially Spain, France, Italy, Greece) with values between 2.0% and 2.7%. The lowest values of $RCV$ (less than 1%) are observed mainly in the eastern Mediterranean Sea. On the other hand, winter is the season with the highest values of $RCV$ (up to 6%), which are depicted in the northern Mediterranean Sea (Italy, Greece, France,

Croatia, Montenegro). The areas with high values during spring are localised at the western coasts of Algeria (around 4%). During summer, the entire basin is characterized by rather low values of *RCV* (below 3%) while in autumn northwestern Aegean, north Adriatic, and Ligurian seas are characterized with values over 4.5%. In general, the southern Aegean and western Alboran seas are characterized by relatively low values of *RCV* for all seasons.

As in the case of WP, *MV* and *SV* are also estimated for the assessment of temporal variability of SP in the monthly and seasonal scales; see Figure 11, where the spatial distribution of *SV* is depicted. Evidently, the lowest values of *SV* are encountered in the African coasts, with values lower than 0.8. The largest values of *SV* (1.2) are located in the northern Adriatic Sea. As regards *MV* (not shown here), it resembles the pattern of *SV*, exhibiting slightly larger values than *SV*.

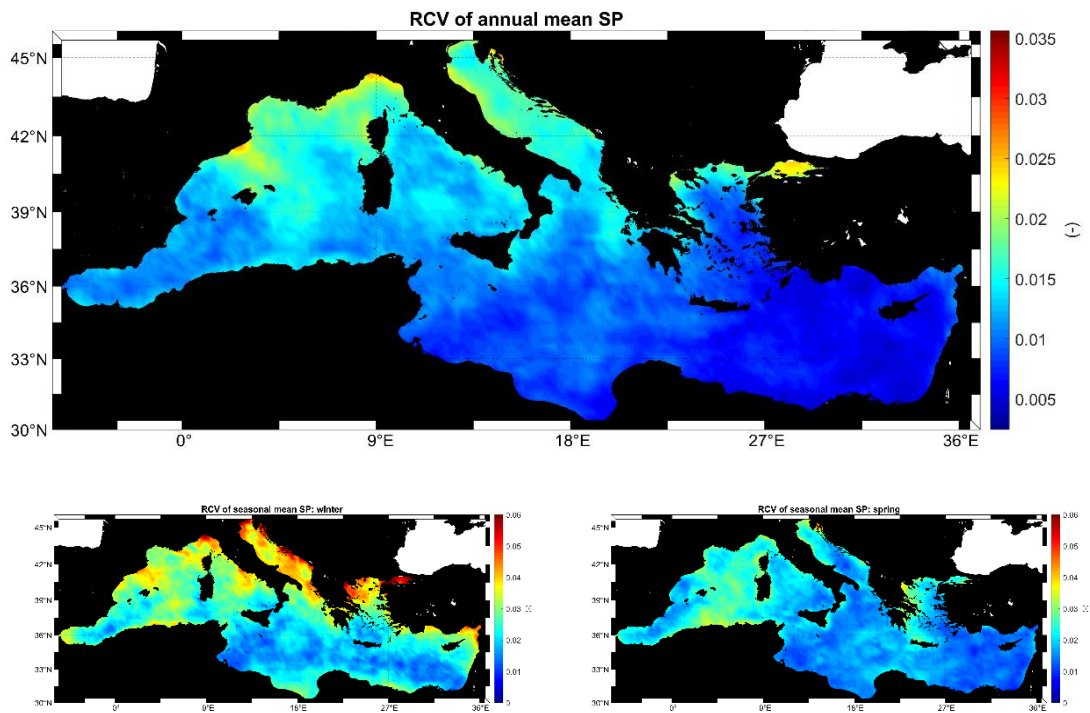

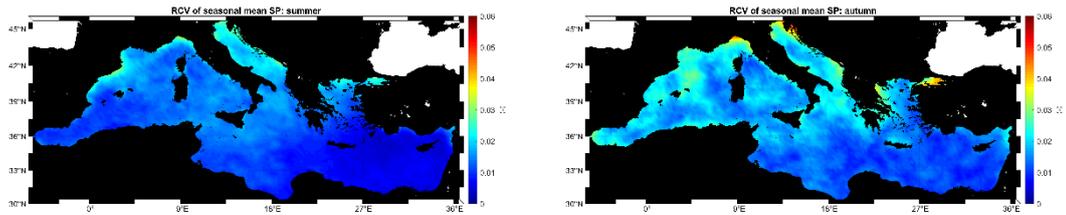

Figure 10. Robust coefficient of variation for the annual (upper panel) and seasonal mean values of solar irradiance for the Mediterranean Sea. Winter: middle left panel, Spring: middle right panel, Summer: lower left panel, Autumn: lower right panel.

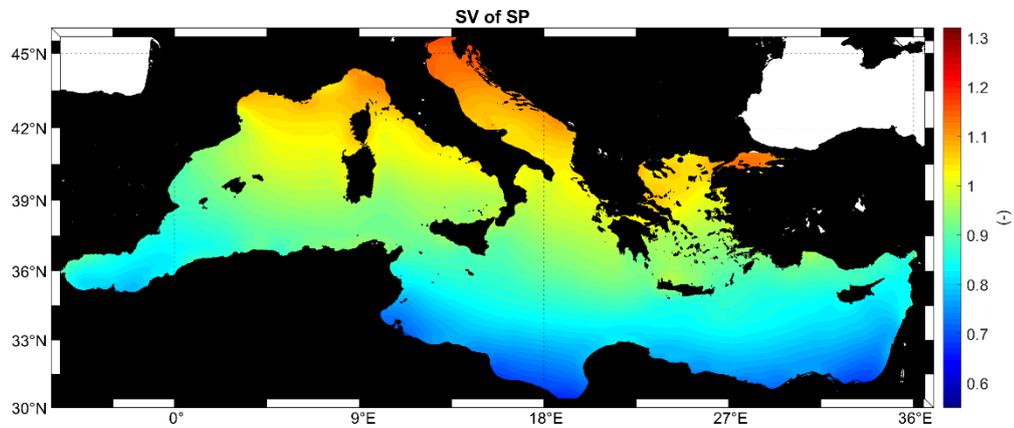

Figure 11. Seasonal variability of solar irradiance for the Mediterranean Sea.

4.3  Joint coefficient of variation of WP and SP

In this section, the joint coefficient of variation $JCV$ of WP and SP for the hourly, seasonal, and annual values is estimated for the Mediterranean Sea. As far as the authors are aware of, $JCV$ is the only metric that describes the joint variability of two energy sources through a unique number (estimate) and is used in applications relevant with hybrid renewable energy sources for the first time.

In Figure 12, $JCV$ for the annual (upper panel) and seasonal mean values of WP and SP is shown. $JCV$ ranges between 0.01 and 0.04 for the annual mean values of WP and SP. The higher values are encountered in the northern coasts of the Mediterranean, while in the southern part, *JCV* remains fairly constant. During winter, the joint variability is more pronounced in the northern areas of the basin (Ligurian and north Adriatic seas, northern Greek coasts, etc.) with values up to 0.08. For spring, summer, and autumn the spatial patterns are very much alike, with the higher values occurring in some localized coastal areas of the northern part (up to 0.05 for spring and autumn). Winter exhibits the largest joint variability, followed by spring, autumn, and summer.

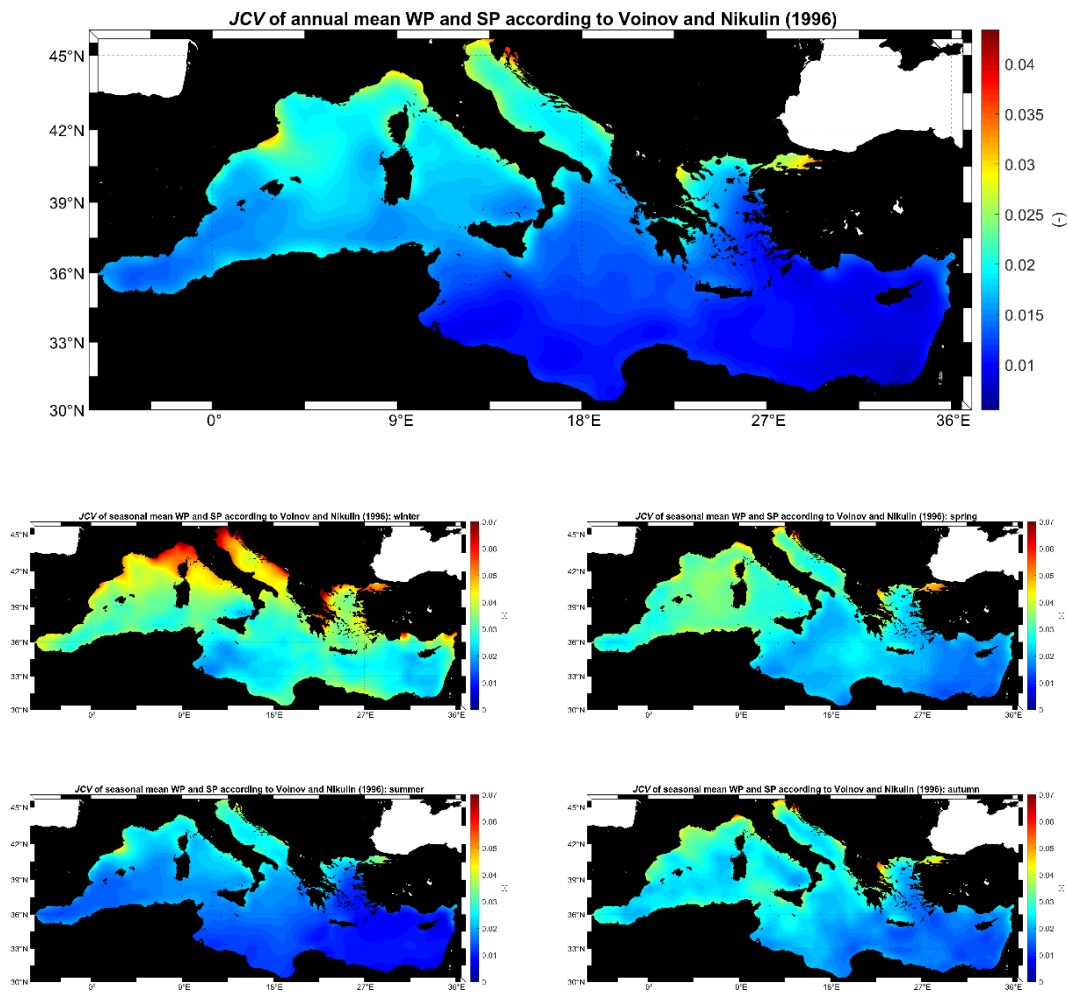

Figure 12. Spatial distribution of the joint coefficient of variation for the annual (upper panel) and seasonal mean values of wind power density and solar irradiance for the Mediterranean

Sea. Winter: middle left panel, Spring: middle right panel, Summer: lower left panel, Autumn: lower right panel.

## 5. COMPLEMENTARITY OF OFFSHORE WIND AND SOLAR RESOURCE

In this section, numerical results are provided for $r_{CMED}$, Pearson's $r$ and Kendall's $\hat{\tau}$ correlation coefficients (CCs), while their comparison is also performed here for the first time at the basin scale. Since CC is a scale value, its interpretation is not easy. In the relevant literature there is no consensus about the qualitative characterization of the strength of the relationship that the value of CC provides. In this work, the following characterizations have been adopted as regards the absolute values of CC: $CC \in (0, 0.2)$ very weak, $CC \in (0.2, 0.4)$ weak, $CC \in (0.4, 0.6)$ moderate, $CC \in (0.6, 0.8)$ strong, and $CC \in (0.8, 1.0)$ very strong. See also [34], [40], [80] for slightly different considerations.

Results for the event-based approach that was presented in section 2.3 are also assessed. The advantage of this approach compared to the correlation-based analysis consists in its immediacy in translating the results in practical terms of complementarity-synergy features between the examined resources.

### 5.1 Correlation based analysis

In Figure 13, the Pearson's $r$ (upper panel), median $r_{CMED}$ (lower left panel), and Kendall's $\hat{\tau}$ (lower right panel) CCs are depicted for the annual mean values of offshore WP and SP. Detailed comments are provided only for the behavior of Pearson's $r$ CC.

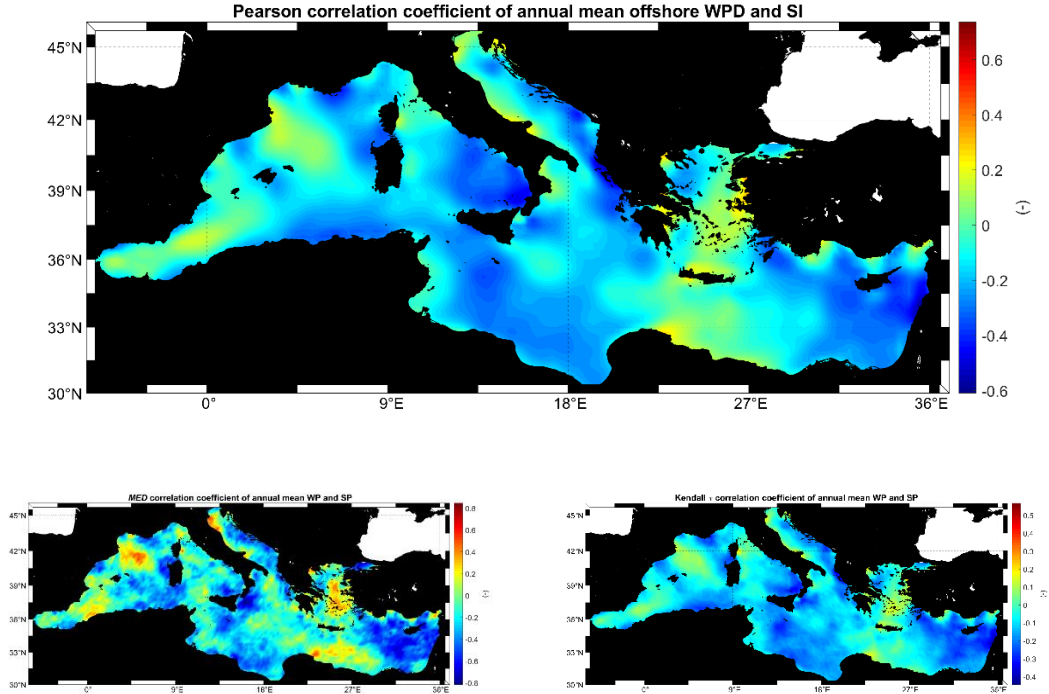

Figure 13. Pearson's $r$ (lower panel), median $r_{CMED}$ (lower left panel), and Kendall's $\hat{\tau}$ (lower right panel) correlation coefficients of the annual mean values of offshore wind power density and solar irradiance for the Mediterranean Sea.

For the majority of the examined water grid points, the values of $r$ are negative. The overall lowest value of $r$ (-0.59) is encountered across the coasts of Lebanon, suggesting a moderate to strong complementarity. Extended areas in the Mediterranean that are characterized by negative values of $r$, are the Levantine, Ligurian, Ionian, and Tyrrhenian Seas. The highest positive value (0.25) is encountered in the coasts of Libya suggesting weak synergy. Extended areas with very low positive values of $r$ (0.1–0.2) are the central and southern Aegean Sea, the area offshore the Gulf of Lion, the area off western Sicily Isl., the eastern coasts of Sardinia and Corsica Isl., the eastern coasts of Libya and the western coasts of Egypt, some localized areas in the southern coasts of Turkey, and the Alboran Sea. Similar spatial patterns are also exhibited by the Kendall's $\hat{\tau}$ and $r_{CMED}$ CCs, but with fairly different absolute values (especially for $r_{CMED}$).

In Table 1, a synoptic comparison of the values provided by the three different CCs is presented. For the overwhelming majority of the total number of grid points in the Mediterranean basin (4,105), the values of $r$ and $\hat{\tau}$ are rather very close. Specifically, for 3,840 grid points (i.e., 93.5% of total points) $r$ and $\hat{\tau}$ provide values with the same direction (sign), while for 3,451 grid points (i.e., 83.1% of total points) the corresponding absolute deviation is less than 0.1 ($|r - \hat{\tau}| \leq 0.1$). On the other hand, $r_{CMED}$ provides a fair number of compatible values with $r$ and $\hat{\tau}$ as regards the corresponding signs, but the number of points with small absolute deviation compared to the other CCs is much smaller.

Table 1: Comparison of the results of $r, r_{CMED}$, and $\hat{\tau}$ for the annual mean values of offshore wind power density and solar irradiance for the Mediterranean Sea.

|  | $r$ | $r_{CMED}$ | $\hat{\tau}$ | $r, r_{CMED}$ | $r, \hat{\tau}$ | $r_{CMED}, \hat{\tau}$ |
|---|---|---|---|---|---|---|
| # of grid points with positive sign | 606 | 829 | 504 | 396 | 419 | 348 |
| # of grid points with negative sign | 3,499 | 3,276 | 3,601 | 3,069 | 3,421 | 3,130 |
| # of grid points with absolute deviation $\leq 0.1$ | - | - | - | 1,674 | 3,451 | 1,542 |

In Figure 14, the Pearson's $r$ (left panels), median $r_{CMED}$ (middle panels), and Kendall's $\hat{\tau}$ (right panels) are depicted for the seasonal mean values of offshore WP and SP.

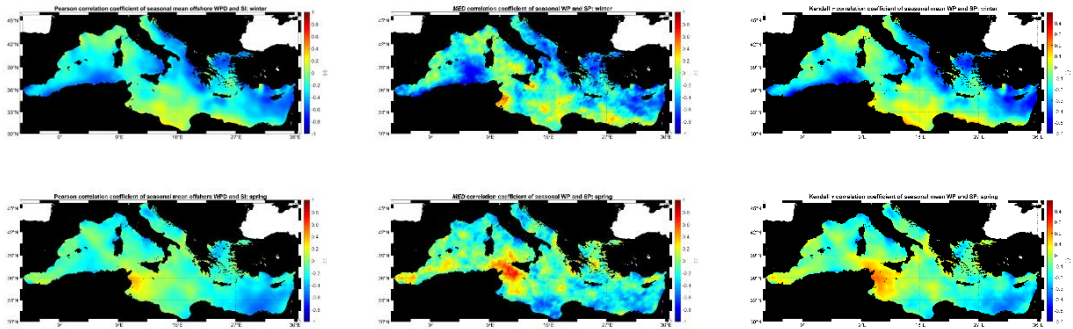

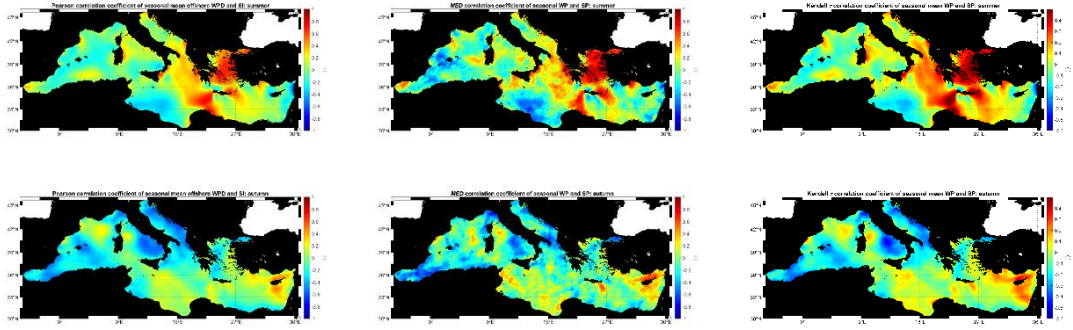

Figure 14. Pearson's $r$ (left panels), median $r_{CMED}$ (middle panels), and Kendall's $\hat{\tau}$ (right panels) correlation coefficients of the seasonal mean values of offshore wind power density and solar irradiance for the Mediterranean Sea. Winter: first row panel, Spring: second row panel, Summer: third row panel, Autumn: fourth row panel.

During winter, extended areas in the Mediterranean exhibit strongly negative values of $r$ especially in the Levantine Sea (-0.72 in the Gulf of Antalya). Other areas characterized by strongly negative values of $r$, are the Ionian (-0.65) and northern Aegean (-0.61) seas, the coasts of Algeria (-0.61), the area off southern Crete Isl., the eastern Adriatic coasts, etc. Weak positive values of $r$ are encountered across the Libyan coasts (up to 0.3). During spring moderate negative values of $r$ are encountered across the coasts of Egypt (-0.59) and moderate positive values in the coasts of Tunisia (0.44). During summer, extended areas exhibit strongly positive values of $r$ reaching up to 0.77, especially in the Aegean (up to 0.72) and Ionian Seas up to African coasts (up to 0.77), and the western coasts of Sardinia and Sicily Isl. Note that all examined CCs reveal the strong association between SP and WP during summer in the Aegean Sea. Moderate and weak negative values of $r$ are encountered in the Alboran Sea (-0.58), offshore Libya (-0.39), and eastern Levantine Sea (-0.34). During autumn, strongly negative values of $r$ are encountered in the northern Adriatic (-0.62), and the Tyrrhenian (-0.6) seas. The highest positive values of $r$ are encountered in the eastern Levantine Sea (0.51) and off southern Crete Isl. (0.25).

In Table 2, a synoptic comparison of the values provided by the three different CCs is presented for each season. For all seasons, the similarity of the results provided by Pearson's $r$ and Kendall's $\hat{\tau}$ CCs is rather evident. The total number of points characterized by the same sign of $r$ and $\hat{\tau}$ exhibits a remarkable stability: 3,898 points for winter; 3,765 points for spring; 3,827 points for summer, and; 3,817 points for autumn. This stability is also present as regards $r, r_{CMED}$ and $\hat{\tau}, r_{CMED}$. The season with the highest number of grid points with positive (negative) sign is summer (winter) for each CC individually and combined. On the other hand, the number of grid points with small absolute deviations ($|r - \hat{\tau}| \leq 0.1$) is: 2,218 for winter; 2,888 for spring; 3,099 for summer, and; 2,702 for autumn. Comparing the results with the corresponding ones of the annual time scale it can be noticed that in the seasonal time scale, the absolute deviation $|r - \hat{\tau}|$ is, for all seasons, greater, while the number of points with the same sign as provided by these CCs is fairly the same. Moreover, the behavior of $r_{CMED}$ compared to the other CCs is almost similar as in the previous case (i.e., for the annual mean values of offshore wind power density and solar irradiance).

Table 2: Comparison of the results of $r, r_{CMED}$, and $\hat{\tau}$ for the seasonal mean values of offshore wind power density and solar irradiance for the Mediterranean Sea.

|  | Season | $r$ | $r_{CMED}$ | $\hat{\tau}$ | $r, r_{CMED}$ | $r, \hat{\tau}$ | $r_{CMED}, \hat{\tau}$ |
|---|---|---|---|---|---|---|---|
| # of grid points with positive sign | Winter | 572 | 816 | 554 | 288 | 455 | 326 |
|  | Spring | 735 | 1,198 | 751 | 437 | 570 | 518 |
|  | Summer | 2,376 | 2,185 | 2,363 | 1,833 | 2,221 | 1,867 |
|  | Autumn | 1,458 | 1,522 | 1,432 | 954 | 1,292 | 972 |
| # of grid points with negative sign | Winter | 3,553 | 3,289 | 3,551 | 3,008 | 3,443 | 3,071 |
|  | Spring | 3,370 | 2,907 | 3,354 | 2,610 | 3,195 | 2,692 |
|  | Summer | 1,729 | 1,920 | 1,742 | 1,385 | 1,606 | 1,445 |
|  | Autumn | 2,647 | 2,583 | 2,673 | 2,089 | 2,525 | 2,147 |
|  | Winter | - | - | - | 1,660 | 2,218 | 1,604 |
| # of grid points with | Spring | - | - | - | 1,604 | 2,888 | 1,691 |
| absolute deviation ≤ | Summer | - | - | - | 1,678 | 3,099 | 1,615 |
| 0.1 | Autumn | - | - | - | 1,713 | 2,702 | 1,855 |

In Figure 15, $r, r_{CMED}$, and $\hat{\tau}$ CCs are depicted for the instantaneous (hourly) values of offshore WP and SP. $r$ is negative for most areas of the Mediterranean Sea; however, the majority of the values (either positive or negative) are very close to 0. In general, in this time scale, there is very weak synergy or very weak complementarity between the resources. Note also that at this time scale, all correlation coefficients are clearly affected by the day-night cycle, [39].

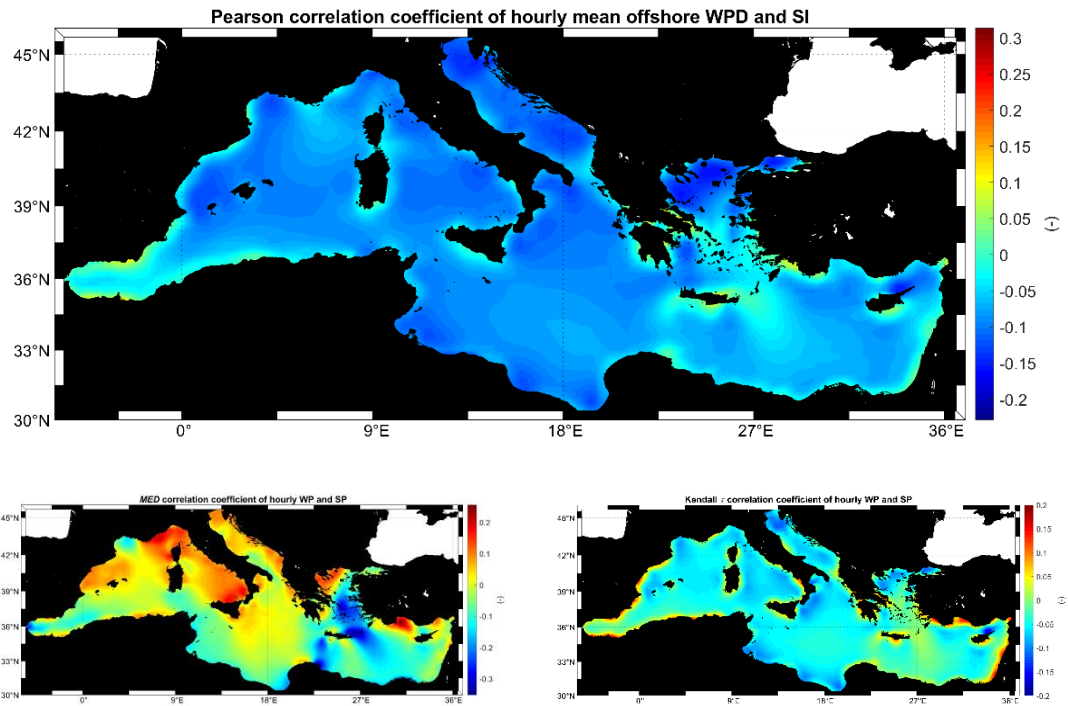

Figure 15. Pearson's $r$ (upper panel), median $r_{CMED}$ (lower left panel), and Kendall's $\hat{\tau}$ (lower right panel) correlation coefficients of hourly values of offshore wind power density and solar irradiance for the Mediterranean Sea.

In Table 3, a synoptic comparison of the values provided by the three different CCs in the hourly time scale is presented. Once more, for the overwhelming majority of the total number of water grid points, the values of $r$ and $\hat{\tau}$ are very close. Particularly, for 3,706 grid points (i.e., for 90.3% of total points), $r$ and $\hat{\tau}$ provide values with the same direction (sign), while for 4,057 grid points (i.e., for 98.8% of total points), the corresponding absolute deviation is less than 0.1 ($|r - \hat{\tau}| \leq 0.1$). On the other hand, $r_{CMED}$ seems again to provide fairly different results as regards both the sign and the corresponding absolute values. Clearly, in the hourly scale, the similarity of the results provided by $r$ and $\hat{\tau}$ is impressive.

Table 3: Comparison of the results of $r, r_{CMED}$, and $\hat{\tau}$ for the hourly values of offshore wind power density and solar irradiance for the Mediterranean Sea

|  | $r$ | $r_{CMED}$ | $\hat{\tau}$ | $r, r_{CMED}$ | $r, \hat{\tau}$ | $r_{CMED}, \hat{\tau}$ |
|---|---|---|---|---|---|---|
| # of grid points with positive sign | 109 | 1,714 | 506 | 30 | 108 | 214 |
| # of grid points with negative sign | 3,996 | 2,391 | 3,599 | 2,312 | 3,598 | 2,099 |
| # of grid points with absolute deviation $\leq 0.1$ | - | - | - | 2,385 | 4,057 | 2,808 |

5.2  Event-based complementarity analysis

The lower thresholds of WP and SP, $WP_L$, and $SP_L$ that were introduced in relations (9), (10), (15), and (16), respectively, are selected as follows: $WP_L$ is considered equal to the upper limit of poor wind power class, i.e., $WP_L \cong 280$ W/m² ([3]); see [57], [43]. Regarding SP, in principle, it can be considered that $SP_L = 0$ W/m² (theoretically a PV module can generate power as long as $SP_L > 0$ W/m²). In this work though, the same approach as for the wind power density is followed, i.e., $SP_L$ is considered equal to the upper limit of poor solar power class, which is 125

---

([3]) Let it be noted that for wind speeds distributed according to the Rayleigh distribution, the following relation holds: $WP_{AN} = \frac{3}{\pi}\rho u_{AN}^3$, where $WP_{AN}$ is the mean annual wind power density and $u_{AN}$ is the mean annual wind speed. For $\rho = 1.225$ kg/m³ and $u_{AN} = 6.2$ m/s, one obtains $WP_{AN} = 280$ W/m².

W/m$^2$. This threshold has been estimated by the National Renewable Energy Laboratory (NREL) ($^4$) based on solar data (measured and modeled) between 1961 and 2008.

In Figure 16, $WCS$ (left panel) and $SCW$ (right panel) are depicted for the Mediterranean Sea. The values of $WCS$ suggest that the entire Aegean and Alboran seas, the extended area off southern Crete Isl., the Gulf of Lion, and the area between Sicily Isl. and Africa are areas where wind complements solar in a moderate degree. According to the values of $SCW$, areas where solar complements wind in a moderate degree are the coasts of middle East, southern Turkey, Greece, Italy and Spain, and the eastern part of the African coasts. Evidently, from a long-term perspective, the solar to wind complementarity is more important, since OWFs are expected to start developing soon in the area, while development of offshore floating solar farms is rather a future consideration. In this context, areas where solar complements wind is the eastern Levantine Sea, the northern coasts of Greece and Sicily Isl., the southern coasts of Italy, and the Mediterranean northern coasts of Spain.

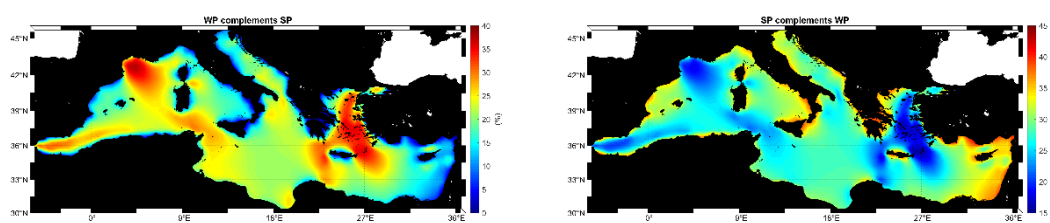

Figure 16. Wind complements solar index ($WCS$) (left panel) and solar complements wind index ($SCW$) (right panel) for the Mediterranean Sea.

In Figure 17, the non-availability index of solar and wind ($USW$, left panel) and the synergy index ($SWS$, right panel) are shown. From a technical point of view, the areas characterized by high values of $USW$ should be excluded from further consideration as regards development of

---

($^4$) See https://openei.org/datasets/dataset/solar-resources-by-class-and-country/resource/3e72f32a-7de1-4e5d-a25a-76928769625f. Date of last access: 1/7/2020.

offshore wind and solar, since neither energy source satisfies the necessary conditions. Areas that exhibit relatively high values of $SWS$, such as the Aegean Sea, the gulfs of Lion, Gabes, and Sidra, and the straits between Sicily Isl. and Africa, are synergetic as regards hybrid development of offshore wind and solar. The highest synergy appears for the gulfs of Lion and Sidra, and the northern Aegean Sea.

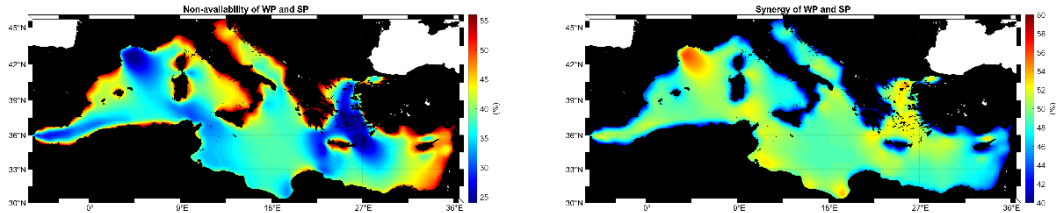

Figure 17. Joint non-availability index (left panel) and synergy index (right panel) for wind and solar for the Mediterranean Sea.

In Figure 18, $\bar{D}_{NW}$ and $\bar{D}_{NS}$ (upper panel), and $\max(D_{NW})$ and $\max(D_{NS})$ (lower panel) are depicted, where $\bar{D}$ denotes the mean value of $D$.

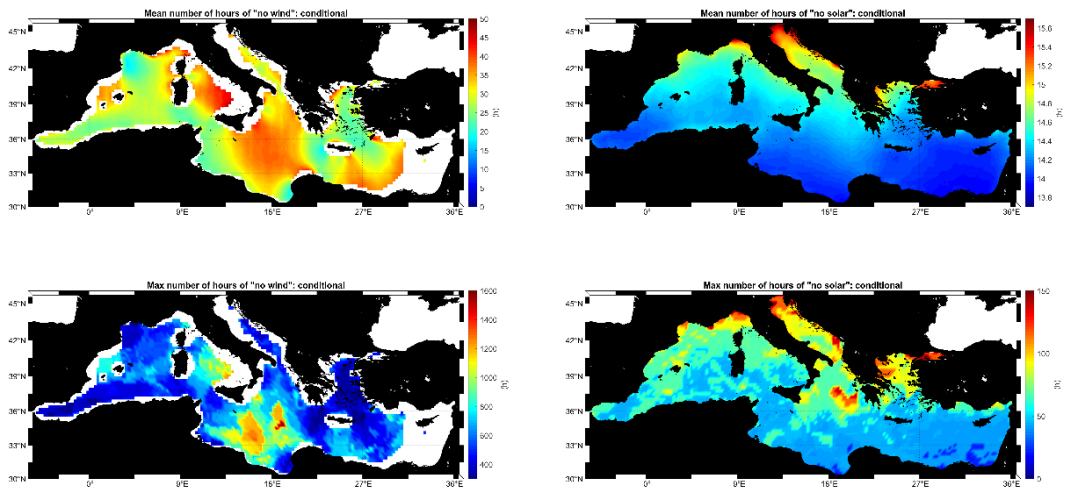

Figure 18. Mean (upper panels) and maximum (lower panels) duration (in hours) of the events $[E_W^C|E_{W,AN}]$ (left panels) and $[E_S^C|E_{S,AN}]$ (right panels) for the Mediterranean Sea.

$\overline{D}_{NW}$ ranges between 15–50 h for the areas that met the values of $WP_L$ on an annual basis. The Tyrrhenian Sea and the area between southern Italy and Africa exhibit the highest values of $\overline{D}_{NW}$ (ranging between 35–45 h). On the other hand, the Aegean and Alboran seas, and the gulfs of Lion, Sidra, and Gabes are characterized by low values of $\overline{D}_{NW}$ (centered at ~25 h), and from this point of view, these areas are potentially more favorable for OWF development. Note that in the areas where $\overline{D}_{NW}$ takes high values, wind energy source is of more intermittent nature, compared to the areas where $\overline{D}_{NW}$ takes lower values. $\overline{D}_{NS}$ ranges between 13.0–14.4 h for the southern Mediterranean coasts that met the values of $SP_L$ on an annual basis. For the northern coasts, $\overline{D}_{NS}$ takes values close to 15.5 h (Venice and Genoa gulfs).

$\max(D_{NW})$ may reach values up to 1,600 h in offshore areas of no particular interest for hybrid offshore wind development (very far from the coast and/or very deep waters). In the Aegean, Adriatic and Alboran seas, as well as in the gulfs of Lion, Sidra, and Gabes, $\max(D_{NW})$ ranges between 300 and 700 h, while in the Balearic Sea it takes slightly larger values. $\max(D_{NS})$ takes lower values (around 40–70 h) in the southern Mediterranean and higher values (around 90–140 h) in the northern part, and especially at the northern coasts of Italy and Greece.

5.3   Resources based on complementarity aspects

In this section, the expected energy output is examined for selected locations in the Mediterranean Sea that exhibit different combinations of the statistical quantities presented beforehand, i.e., power potential, variability, correlation, complementarity, and availability. For each location, the energy output is estimated for each energy farm separately and jointly, in case a hybrid application is considered. In this context, six locations along the Mediterranean basin are selected with diverse behavior among each other. The details for each considered measure is summarized in Table 4. Ideal locations with high both wind and solar energy potential and high complementarity are absent in the Mediterranean Sea, thus the selected sites highlight regimes that could favor either the development of a hybrid plant or each installation individually. The qualitative characterization for all quantities, except for $r$, includes the following classes: very low, low, medium, high, very high. The classes for $r$ were provided in section 5. For instance, Alboran Sea exhibits high wind and solar power (with low and very low variability, respectively) and medium synergy for the two sources while Gulf of Lion presents a high synergetic behavior, but with high intermittency for solar resource, and very high wind power potential and low solar power.

Table 4: Geographical coordinates and qualitative characterization of the statistical quantities considered for each examined location (VH: very high, H: high, M: medium/moderate, W: weak, L: low, VL: very low, ~0: almost zero, VW: very weak, +: positive, -: negative).

| Name | Location (lat, lon) | WP, MAV | SP, MAV | $r$ | SWS | $\overline{D}_{NS}$ |
|---|---|---|---|---|---|---|
| Alboran Sea | (36.25, -3) | H, L | H, VL | ~0 | M | L |
| Central Aegean Sea | (37.5, 25) | H, VL | M, M | VW, + | M | L |
| Gulf of Lion | (42.75, 3.5) | VH, L | L, M | ~0 | H | H |
| Gulf of Sidra | (30.5, 19.5) | L, L | VH, VL | W - | H | VL |
| North-eastern Sicily Isl. | (38.75, 15.5) | L, M | L, M | M - | L | M |
| Southern Cyprus Isl. | (34.25, 33.25) | L, L | H, VL | W - | M | VL |

For the exploitation of offshore wind energy, an 8-MW offshore wind turbine was chosen, the Vestas V164-8.0 produced by MHI Vestas Offshore Wind A/S, Denmark, specifically designed for offshore wind conditions. The power curve and the technical specifications are presented in Table 5. The mean power output of a wind turbine $\overline{P}_W$ at a specific location can be estimated by combining its power curve and the available wind speed time series (extrapolated to the turbine hub height) as follows:

$$\bar{P}_W = \frac{1}{N}\sum_{i=1}^{N} P_W(u_i), \qquad (19)$$

where $N$ is the sample size. The monthly estimates of the above quantity can be estimated in a straightforward way. The corresponding energy output $E_W$ can be obtained by integrating $P_W$ over each month.

Table 5: Power curve of the Vestas V164-8.0 wind turbine along with the technical specifications. (Source: https://en.wind-turbine-models.com)

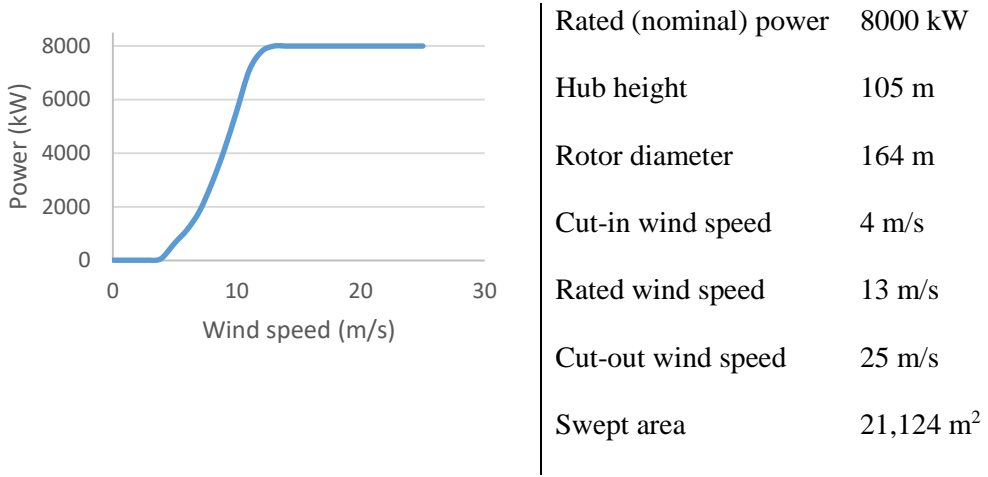

| | |
|---|---|
| Rated (nominal) power | 8000 kW |
| Hub height | 105 m |
| Rotor diameter | 164 m |
| Cut-in wind speed | 4 m/s |
| Rated wind speed | 13 m/s |
| Cut-out wind speed | 25 m/s |
| Swept area | 21,124 m² |

In order to perform a meaningful assessment of the energy output from the two resources, the same capacity (i.e., 8 MW) was considered for solar energy, as proposed by [27] and the corresponding floating solar farm is considered. The SPR-220-BLK solar panel developed by SunPower was chosen along with the Schneider Electric GT500-480 [480V] solar inverter [81]. The details for their technical characteristics are shown in Table 6. Based on the nominal power of the solar panel and the total nominal power considered, the total number of solar PV panels ($N_{PV}$) required is 36,364. According to the dimensions of each solar panel, the total sea surface area that is covered by this solar farm is 45,237 m², which approximately equals to 212.7 m x 212.7 m (i.e., 171 solar panels for $x-, y-$direction). Bearing in mind the co-exploitation of wind and sun, if a minimum distance of 3 rotor diameters is considered between the wind turbines in the direction perpendicular to dominant wind direction [82], the "free" sea surface is more than sufficient to host a solar system of this size.

Table 6: The technical specifications of the floating PV system considered in this study.

| Solar panel: SPR-220-BLK | Solar inverter: Schneider Electric GT500-480 |
|---|---|
| Rated Power: 220 W | Rated Power: AC: 500 kW |
| Nominal Efficiency: 17.8% | Input voltage range, MPPT: 310 - 480 V |
| Temperature coefficient:-0.041/°C | Output voltage: 480 V |
| Voltage Module: 40.8 V | Inverter DC to AC ratio: 0.99 |
| Length x Width: 1.244 m x 1 m | Efficiency: 97% |

The generated temperature-corrected power $P_S$ of a given solar panel is given by [81], [83]:

$$P_S = \eta A G' P_{STC}[1 - \alpha_T(T_{mod} - T_{STC})], \qquad (20)$$

where $\eta$ is a performance factor that denotes the overall efficiency of the system, which takes into account various factors that reduce the efficiency of the installation (e.g., shading, losses, soiling of panels), $G' = G/G_{STC}$ with $G$ denoting the solar irradiance and $G_{STC} = 1,000$ W/m² the solar irradiance at standard testing conditions (STC), $P_{STC}$ is the rated power of the PV module, $\alpha_T$ is the temperature coefficient of the PV module, $T_{mod}$ is the PV module temperature, and $T_{STC} = 25$ °C is the reference solar PV temperature at STC, and $A$ the panel area. The efficiency $\eta$ of the system is considered in this study equal to 0.85, [27]. Under specific weather conditions and taking into consideration the cooling effect of the water, $T_{mod}$ can be estimated empirically by the following function:

$$T_{mod} = c_0 + c_1 T_a + c_2 G - c_3 u_{10}, \qquad (21)$$

where $c_0 = 2.0458$ °C, $c_1 = 0.9458$ °C$^{-1}$, $c_2 = 0.0215$ °Cm²/W, $c_3 = 1.2376$ °Cs/m are empirical coefficients, $T_a$ is the ambient temperature, and $u_{10}$ is the wind speed at 10 m above sea level; see [84]. The energy output $E_S$ can be obtained in a similar way with $E_W$.

In Figure 19, the results of the monthly values of energy output for the offshore wind turbine, the floating solar farm and their combination are presented at the selected locations. The offshore wind energy output takes values greater than 2.5 GWh for Gulf of Lion and Aegean and Alboran seas during all months (apart from August-October at Alboran Sea and May at Aegean Sea). Alboran Sea seems to have a rather stable offshore wind energy production throughout the year while the opposite is valid for northeastern Sicily Isl. In Gulf of Sidra, southern Cyprus Isl., and northeastern Sicily Isl., solar energy production exceeds offshore wind energy production for the periods June-October (5 months), May-October (6 months), and May-September (5 months), respectively, with the most pronounced differences found at the latter location. As expected, the values of energy output for January, February, November, and December are the lowest (smaller than 1 GWh) for all locations regarding the solar resource. In case of a hybrid plant, the Aegean Sea seems to be the most favourable location, reaching 6 GWh of total energy output for July and August, two months with high energy demand as well, resulting mainly from the high wind speeds during this period of the year. Alboran Sea (up to 4.9 GWh for May) and Gulf of Lion (up to 4.7 GWh for July) are the next favourable sites. Despite the low energy output at northeastern Sicily Isl. compared to the rest examined locations, the combined exploitation of offshore wind and solar irradiance seems to be a meaningful choice as it results to a more stable energy output during the year.

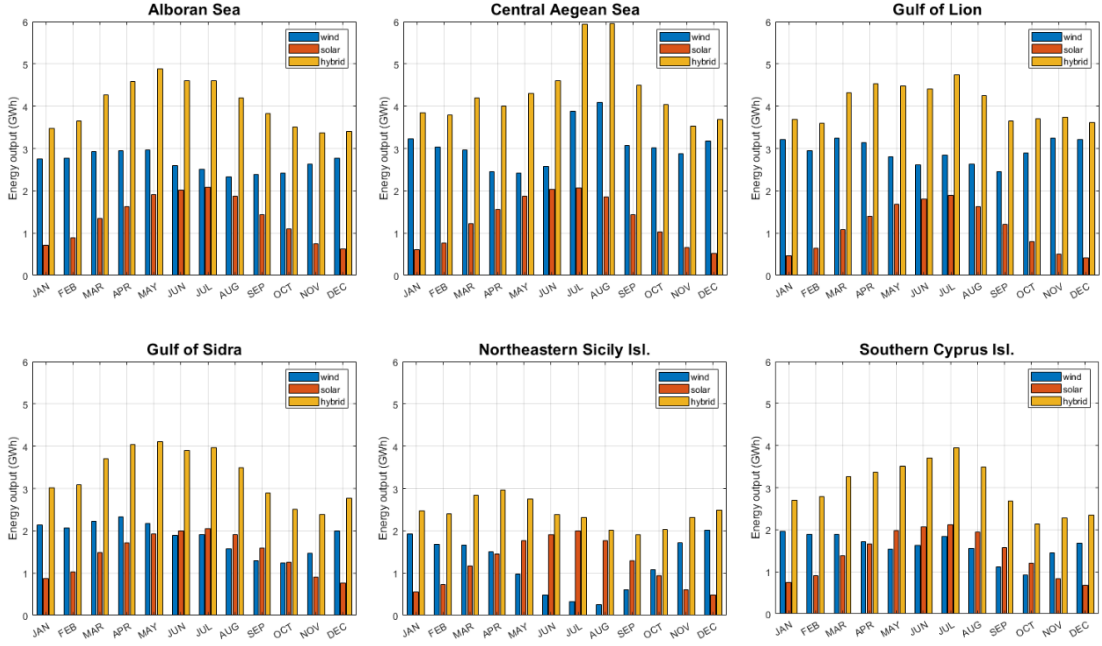

Figure 19: Monthly values of energy output of the 8-MW offshore wind turbine, the 8-MW floating solar farm, and their combination at Alboran Sea (upper left), central Aegean Sea (upper middle), Gulf of Lion (upper right), Gulf of Sidra (lower left), northeastern Sicily Isl. (lower middle), and southern Cyprus Isl. (lower right).

## 6. CONCLUSIONS

A detailed analysis of the offshore wind and solar energy potential was performed for the Mediterranean Sea and several relevant aspects were assessed in depth. The energy resources were examined both independently (univariate analysis) and jointly. Apart from $MAV$ and $IAV$ measures, the spatiotemporal variability was also assessed by implementing the robust coefficient of variation $RCV$, since $CV$ suffers from oversensitivity for distributions with long tails. The robust correlation coefficient $r_{CMED}$, along with Kendall's $\tau$ and Pearson's $r$, were applied for the joint assessment of the two examined marine resources. Moreover, a rather unknown measure for the joint variability of WP and SP, the joint coefficient of variation $JCV$, was introduced and applied for the first time at the seasonal and annual scale. The main conclusions regarding power potential, variability, correlation, and complementarity between WP and SP are summarized below:

- Aegean and Alboran seas, and the gulfs of Lion, Sidra, and Gabes are promising areas for OWF development due to the high offshore wind potential and their low variability.
- Eastern Mediterranean, gulfs of Sidra and Gabes, and Alboran Sea are characterized by relatively low values of $MAV$, $IAV$, and $RCV$, and high SP values, rendering these regions promising for floating solar farm development.
- From the joint assessment in terms of CCs, the results reveal that the central Aegean Sea, offshore the Gulf of Lion and the eastern coasts of Libya are characterized by high positive values, while the negative values are encountered in the eastern Mediterranean, southern Italy, offshore western Sardinia Isl., and the east Adriatic coasts. In terms of joint variability, the areas from the 38$^{th}$ parallel north and further south present low values of $JCV$.
- Wind complements solar in a high degree in the Aegean and Alboran seas and the Gulf of Lion, while solar complements wind in the eastern Levantine Sea and the coastal areas of Greece, Italy and Spain. Nevertheless, the latter areas normally should not be considered for further analysis since the examined energy resources are simultaneously unavailable for extended periods.
- When the combination of temporal persistence and sufficient resource availability is of interest, the Tyrrhenian Sea and the area between southern Italy and Africa are most promising for WP while the northern coasts of Italy and Greece are favourable for SP.

The performance of an offshore wind turbine and a floating solar farm of the same rated power was also assessed for six particular locations in the examined basin. The results showed that the consideration of combining offshore wind and floating solar plants is substantial even for medium degree of complementarity. The complementarity analysis presented in this work can be further extended and focus on additional technical criteria and environmental restrictions.

This would significantly contribute to the actual identification of particular favourable locations in the Mediterranean Sea for hybrid offshore wind and solar farm future developments. Moreover, since climate change is expected to have impacts on renewable energy resources (in terms of supply and demand), it is also crucial to give priority to studies related with the effects of climate change on the complementarity of offshore wind and solar energy for a number of different scenarios.

**Acknowledgements**: ERA5 data is available on the Copernicus Climate Change Service (C3S) Climate Data Store: https://cds.climate.copernicus.eu/#!/search?text=ERA5&type=dataset.

**Funding**: Flora Karathanasi was partially supported for this work, which is part of a project that has received funding from the European Union's Horizon 2020 research and innovation programme under grant agreement No 857586.

## 7. REFERENCES

[1] A.N. Penna. A History of Energy Flows: From Human Labor to Renewable Power. Taylor & Francis2019.

[2] A. Lavagnini, A.M. Sempreviva, C. Transerici, C. Accadia, M. Casaioli, S. Mariani, et al. Offshore wind climatology over the Mediterranean basin. Wind Energy. 9 (2006) 251-66.

[3] M. Menendez, M. Garcia-Diez, L. Fita, J. Fernandez, F.J. Mendez, J.M. Gutierrez. High-resolution sea wind hindcasts over the Mediterranean area. Clim Dynam. 42 (2014) 1857-72.

[4] I. Balog, P.M. Ruti, I. Tobin, V. Armenio, R. Vautard. A numerical approach for planning offshore wind farms from regional to local scales over the Mediterranean. Renew Energ. 85 (2016) 395-405.

[5] T. Soukissian, F. Karathanasi, P. Axaopoulos. Satellite-Based Offshore Wind Resource Assessment in the Mediterranean Sea. Ieee J Oceanic Eng. 42 (2017) 73-86.

[6] D. Pantusa, G.R. Tomasicchio. Large-scale offshore wind production in the Mediterranean Sea. Cogent Eng. 6 (2019).

[7] M.M. Nezhad, D. Groppi, P. Marzialetti, L. Fusilli, G. Laneve, F. Cumo, et al. Wind energy potential analysis using Sentinel-1 satellite: A review and a case study on Mediterranean islands. Renew Sust Energ Rev. 109 (2019) 499-513.

[8] F. Ferrari, G. Besio, F. Cassola, A. Mazzino. Optimized wind and wave energy resource assessment and offshore exploitability in the Mediterranean Sea. Energy. 190 (2020).

[9] T.H. Soukissian, A. Papadopoulos. Effects of different wind data sources in offshore wind power assessment. Renew Energ. 77 (2015) 101-14.

[10] I. Koletsis, V. Kotroni, K. Lagouvardos, T. Soukissian. Assessment of offshore wind speed and power potential over the Mediterranean and the Black Seas under future climate changes. Renew Sust Energ Rev. 60 (2016) 234-45.

[11] E. Skoplaki, J.A. Palyvos. On the temperature dependence of photovoltaic module electrical performance: A review of efficiency/power correlations. Sol Energy. 83 (2009) 614-24.

[12] L.X. Zhu, A.P. Raman, S.H. Fan. Radiative cooling of solar absorbers using a visibly transparent photonic crystal thermal blackbody. P Natl Acad Sci USA. 112 (2015) 12282-7.

[13] A. Sahu, N. Yadav, K. Sudhakar. Floating photovoltaic power plant: A review. Renewable and Sustainable Energy Reviews. 66 (2016) 815-24.

[14] C. Diendorfer, M. Haider, M. Lauermann. Performance Analysis of Offshore Solar Power Plants. Energy Procedia. 49 (2014) 2462-71.

[15] K. Trapani, D.L. Millar. Proposing offshore photovoltaic (PV) technology to the energy mix of the Maltese islands. Energ Convers Manage. 67 (2013) 18-26.

[16] C. Solanki, G. Nagababu, S. Kachhwah. Assessment of offshore solar energy along the coast of India. Energy Procedia. 138 (2017) 530-5.

[17] Y.N. Wu, L.W.Y. Li, Z.X. Song, X.S. Lin. Risk assessment on offshore photovoltaic power generation projects in China based on a fuzzy analysis framework. J Clean Prod. 215 (2019) 46-62.

[18] J. Gao, F. Guo, X. Li, X. Huang, H. Men. Risk assessment of offshore photovoltaic projects under probabilistic linguistic environment. Renew Energ. 163 (2021) 172-87.

[19] C. Kalogeri, G. Galanis, C. Spyrou, D. Diamantis, F. Baladima, M. Koukoula, et al. Assessing the European offshore wind and wave energy resource for combined exploitation. Renew Energ. 101 (2017) 244-64.

[20] A. Azzellino, C. Lanfredi, L. Riefolo, V. De Santis, P. Contestabile, D. Vicinanza. Combined Exploitation of Offshore Wind and Wave Energy in the Italian Seas: A Spatial Planning Approach. Front Energy Res. 7 (2019).

[21] S. Astariz, G. Iglesias. Selecting optimum locations for co-located wave and wind energy farms. Part I: The Co-Location Feasibility index. Energ Convers Manage. 122 (2016) 589-98.

[22] S. Astariz, G. Iglesias. Selecting optimum locations for co-located wave and wind energy farms. Part II: A case study. Energ Convers Manage. 122 (2016) 599-608.

[23] S. Astariz, G. Iglesias. Enhancing Wave Energy Competitiveness through Co-Located Wind and Wave Energy Farms. A Review on the Shadow Effect. Energies. 8 (2015) 7344-66.

[24] S. Astariz, G. Iglesias. Output power smoothing and reduced downtime period by combined wind and wave energy farms. Energy. 97 (2016) 69-81.

[25] A. Hasan, I. Dincer. A new performance assessment methodology of bifacial photovoltaic solar panels for offshore applications. Energ Convers Manage. 220 (2020) 112972.

[26] N.M. Kumar. Model to estimate the potential and performance of Wavevoltaics. Results Phys. 12 (2019) 914-6.

[27] M. López, N. Rodríguez, G. Iglesias. Combined Floating Offshore Wind and Solar PV. J Mar Sci Eng. 8 (2020) 576.

[28] S. Zereshkian, D. Mansoury. A study on the feasibility of using solar radiation energy and ocean thermal energy conversion to supply electricity for offshore oil and gas fields in the Caspian Sea. Renew Energ. 163 (2021) 66-77.

[29] S. Oliveira-Pinto, P. Rosa-Santos, F. Taveira-Pinto. Assessment of the potential of combining wave and solar energy resources to power supply worldwide offshore oil and gas platforms. Energ Convers Manage. 223 (2020) 113299.

[30] J. Jurasz, F.A. Canales, A. Kies, M. Guezgouz, A. Beluco. A review on the complementarity of renewable energy sources: Concept, metrics, application and future research directions. Sol Energy. 195 (2020) 703-24.

[31] L. Xu, Z. Wang, Y. Liu. The spatial and temporal variation features of wind-sun complementarity in China. Energ Convers Manage. 154 (2017) 138-48.

[32] S. Han, L.-n. Zhang, Y.-q. Liu, H. Zhang, J. Yan, L. Li, et al. Quantitative evaluation method for the complementarity of wind–solar–hydro power and optimization of wind–solar ratio. Appl Energ. 236 (2019) 973-84.

[33] M. Denault, D. Dupuis, S. Couture-Cardinal. Complementarity of hydro and wind power: Improving the risk profile of energy inflows. Energy Policy. 37 (2009) 5376-84.

[34] M.P. Cantão, M.R. Bessa, R. Bettega, D.H.M. Detzel, J.M. Lima. Evaluation of hydro-wind complementarity in the Brazilian territory by means of correlation maps. Renew Energ. 101 (2017) 1215-25.

[35] A. Beluco, P.K. de Souza, A. Krenzinger. A dimensionless index evaluating the time complementarity between solar and hydraulic energies. Renew Energ. 33 (2008) 2157-65.

[36] L. Yi, V.G. Agelidis, Y. Shrivastava. Wind-solar resource complementarity and its combined correlation with electricity load demand. 2009 4th IEEE Conference on Industrial Electronics and Applications2009. pp. 3623-8.

[37] J. Widen. Correlations Between Large-Scale Solar and Wind Power in a Future Scenario for Sweden. Ieee T Sustain Energ. 2 (2011) 177-84.

[38] S. Jerez, R.M. Trigo, A. Sarsa, R. Lorente-Plazas, D. Pozo-Vázquez, J.P. Montávez, et al. Spatio-temporal Complementarity between Solar and Wind Power in the Iberian Peninsula. Energy Procedia. 40 (2013) 48-57.

[39] F. Monforti, T. Huld, K. Bodis, L. Vitali, M. D'Isidoro, R. Lacal-Arantegui. Assessing complementarity of wind and solar resources for energy production in Italy. A Monte Carlo approach. Renew Energ. 63 (2014) 576-86.

[40] M.M. Miglietta, T. Huld, F. Monforti-Ferrario. Local Complementarity of Wind and Solar Energy Resources over Europe: An Assessment Study from a Meteorological Perspective. Journal of Applied Meteorology and Climatology. 56 (2017) 217-34.

[41] W. Zappa, M. van den Broek. Analysing the potential of integrating wind and solar power in Europe using spatial optimisation under various scenarios. Renew Sust Energ Rev. 94 (2018) 1192-216.

[42] M.A. Vega-Sánchez, P.D. Castañeda-Jiménez, R. Peña-Gallardo, A. Ruiz-Alonso, J.A. Morales-Saldaña, E.R. Palacios-Hernández. Evaluation of complementarity of wind and solar energy resources over Mexico using an image processing approach. 2017 IEEE International Autumn Meeting on Power, Electronics and Computing (ROPEC)2017. pp. 1-5.

[43] A.A. Prasad, R.A. Taylor, M. Kay. Assessment of solar and wind resource synergy in Australia. Appl Energ. 190 (2017) 354-67.

[44] G. Ren, J. Wan, J. Liu, D. Yu. Spatial and temporal assessments of complementarity for renewable energy resources in China. Energy. 177 (2019) 262-75.

[45] D. Schindler, H.D. Behr, C. Jung. On the spatiotemporal variability and potential of complementarity of wind and solar resources. Energ Convers Manage. 218 (2020) 113016.

[46] S. Sterl, S. Liersch, H. Koch, N.P.M.v. Lipzig, W. Thiery. A new approach for assessing synergies of solar and wind power: implications for West Africa. Environmental Research Letters. 13 (2018) 094009.

[47] F. Weschenfelder, G.D.P. Leite, A.C.A. da Costa, O.D. Vilela, C.M. Ribeiro, A.A.V. Ochoa, et al. A review on the complementarity between grid-connected solar and wind power systems. J Clean Prod. 257 (2020).

[48] F.J. Santos-Alamillos, D. Pozo-Vazquez, J.A. Ruiz-Arias, V. Lara-Fanego, J. Tovar-Pescador. A methodology for evaluating the spatial variability of wind energy resources: Application to assess the potential contribution of wind energy to baseload power. Renew Energ. 69 (2014) 147-56.

[49] I. Ashton, J.C.C. Van-Nieuwkoop-McCall, H.C.M. Smith, L. Johanning. Spatial variability of waves within a marine energy site using in-situ measurements and a high resolution spectral wave model. Energy. 66 (2014) 699-710.

[50] T. Soukissian, F. Karathanasi, P. Axaopoulos, E. Voukouvalas, V. Kotroni. Offshore wind climate analysis and variability in the Mediterranean Sea. Int J Climatol. 38 (2018) 384-402.

[51] S.M. Fisher, J.T. Schoof, C.L. Lant, M.D. Therrell. The effects of geographical distribution on the reliability of wind energy. Applied Geography. 40 (2013) 83-9.

[52] S.J. Watson. Quantifying the Variability of Wind Energy. WIRES Energ Environ. 3 (2014) 330-42.

[53] K. Engeland, M. Borga, J.-D. Creutin, B. François, M.-H. Ramos, J.-P. Vidal. Space-time variability of climate variables and intermittent renewable electricity production – A review. Renewable and Sustainable Energy Reviews. 79 (2017) 600-17.

[54] J.C.Y. Lee, M.J. Fields, J.K. Lundquist. Assessing variability of wind speed: comparison and validation of 27 methodologies. Wind Energ Sci. 3 (2018) 845-68.

[55] R. Ospina, F. Marmolejo-Ramos. Performance of Some Estimators of Relative Variability. Frontiers in Applied Mathematics and Statistics. 5 (2019).

[56] C.N.P.G. Arachchige, L.A. Prendergast, R.G. Staudte. Robust analogs to the coefficient of variation. Journal of Applied Statistics. (2020) 1-23.

[57] U.B. Gunturu, C.A. Schlosser. Characterization of wind power resource in the United States. Atmos Chem Phys. 12 (2012) 9687-702.

[58] A. Cornett. A global wave energy assessment. The Eighteenth International Offshore and Polar Engineering Conference. International Society of Offshore and Polar Engineers, Vancouver, Canada, 2008. p. 9.

[59] S. Aerts, G. Haesbroeck, C. Ruwet. Multivariate coefficients of variation: Comparison and influence functions. J Multivariate Anal. 142 (2015) 183-98.

[60] V.G. Voinov, M.S. Nikulin. Unbiased Estimators and their Applications. Volume 2: Multivariate Case. 1st ed. Springer Netherlands1996.

[61] G. Shevlyakov, P. Smirnov. Robust Estimation of the Correlation Coefficient: An Attempt of Survey. Austrian Journal of Statistics. 40 (2016) 147–56.

[62] Z.V. Li, G.L. Shevlyakov, V.I. Shin. Robust estimation of a correlation coefficient for ε-contaminated bivariate normal distributions. Automat Rem Contr+. 67 (2006) 1940-57.

[63] G.L. Shevlyakov. On Robust estimation of a correlation coefficient. Journal of Mathematical Sciences. 83 (1997) 434-8.

[64] H. Hersbach, D. Dee. ERA5 reanalysis is in production, . Available at: https://www.ecmwf.int/en/newsletter/147/news/era5-reanalysis-production (last access: 14 November 2018), ECMWF Newsletter, Vol. 147, p. 7, 2016.

[65] Copernicus Climate Change Service. ERA5: Fifth generation of ECMWF atmospheric reanalyses of the global climate. in: Copernicus Climate Change Service Climate Data Store (CDS), (Ed.).2017.

[66] H. Farjami, A.R.E. Hesari. Assessment of sea surface wind field pattern over the Caspian Sea using EOF analysis. Reg Stud Mar Sci. 35 (2020).

[67] L.F.D. Tavares, M. Shadman, L.P.D. Assad, C. Silva, L. Landau, S.F. Estefen. Assessment of the offshore wind technical potential for the Brazilian Southeast and South regions. Energy. 196 (2020).

[68] A. Ulazia, J. Saenz, G. Ibarra-Berastegi, S.J. Gonzalez-Roji, S. Carreno-Madinabeitia. Global estimations of wind energy potential considering seasonal air density changes. Energy. 187 (2019).

[69] J. Olauson. ERA5: The new champion of wind power modelling? Renew Energ. 126 (2018) 322-31.

[70] S. Aniskevich, V. Bezrukovs, U. Zandovskis, D. Bezrukovs. Modelling the Spatial Distribution of Wind Energy Resources in Latvia. Latvian Journal of Physics and Technical Sciences. 54 (2017) 10-20.

[71] D. Allaerts, S.V. Broucke, N. van Lipzig, J. Meyers. Annual impact of wind-farm gravity waves on the Belgian–Dutch offshore wind-farm cluster. Journal of Physics: Conference Series. 1037 (2018) 072006.

[72] P. Bechtle, M. Schelbergen, R. Schmehl, U. Zillmann, S. Watson. Airborne wind energy resource analysis. Renew Energ. 141 (2019) 1103-16.

[73] R. Urraca, T. Huld, A. Gracia-Amillo, F.J. Martinez-de-Pison, F. Kaspar, A. Sanz-Garcia. Evaluation of global horizontal irradiance estimates from ERA5 and COSMO-REA6 reanalyses using ground and satellite-based data. Sol Energy. 164 (2018) 339-54.

[74] H. Jiang, Y.P. Yang, H.Z. Wang, Y.Q. Bai, Y. Bai. Surface Diffuse Solar Radiation Determined by Reanalysis and Satellite over East Asia: Evaluation and Comparison. Remote Sens-Basel. 12 (2020).

[75] D. Yang, J.M. Bright. Worldwide validation of 8 satellite-derived and reanalysis solar radiation products: A preliminary evaluation and overall metrics for hourly data over 27 years. Sol Energy. (2020) in press.

[76] B. Yao, C. Liu, Y. Yin, Z.Q. Liu, C.X. Shi, H. Iwabuchi, et al. Evaluation of cloud properties from reanalyses over East Asia with a radiance-based approach. Atmos Meas Tech. 13 (2020) 1033-49.

[77] J. Huang, L.J. Rikus, Y. Qin, J. Katzfey. Assessing model performance of daily solar irradiance forecasts over Australia. Sol Energy. 176 (2018) 615-26.

[78] B. Babar, R. Graversen, T. Bostrom. Solar radiation estimation at high latitudes: Assessment of the CMSAF databases, ASR and ERA5. Sol Energy. 182 (2019) 397-411.

[79] M. De Felice, M.B. Soares, A. Alessandri, A. Troccoli. Scoping the potential usefulness of seasonal climate forecasts for solar power management. Renew Energ. 142 (2019) 215-23.

[80] A.H. Chughtai, N.U. Hassan, C. Yuen. Planning for Mitigation of Variability in Renewable Energy Resources using Temporal Complementarity. 2018 IEEE Innovative Smart Grid Technologies - Asia (ISGT Asia)2018. pp. 1147-52.

[81] A.P. Sukarso, K.N. Kim. Cooling Effect on the Floating Solar PV: Performance and Economic Analysis on the Case of West Java Province in Indonesia. Energies. 13 (2020) 2126.

[82] P. Hou, J. Zhu, K. Ma, G. Yang, W. Hu, Z. Chen. A review of offshore wind farm layout optimization and electrical system design methods. Journal of Modern Power Systems and Clean Energy. 7 (2019) 975-86.

[83] T. Huld, M. Šúri, E.D. Dunlop. Geographical variation of the conversion efficiency of crystalline silicon photovoltaic modules in Europe. Progress in Photovoltaics: Research and Applications. 16 (2008) 595-607.

[84] W. Charles Lawrence Kamuyu, J.R. Lim, C.S. Won, H.K. Ahn. Prediction Model of Photovoltaic Module Temperature for Power Performance of Floating PVs. Energies. 11 (2018) 447.